\documentclass[superscriptaddress,showpacs,nofootinbib,notitlepage,preprintnumbers,secnumarabic,amssymb, nobibnotes,aps, prd,twocolumn,10pt]{revtex4-1}
\usepackage[utf8]{inputenc}
\usepackage{graphicx}
\usepackage{latexsym,amsmath,amssymb,amsthm,lmodern,float,url,bm}
\usepackage{natbib}
\usepackage{color}
\usepackage{microtype}
\usepackage{import}
\usepackage{bbold}
\usepackage[plain]{fancyref}
\usepackage{varioref}
\usepackage{slashed}
\usepackage{multirow}
\usepackage{tikz}
\usepackage{physics}
\usepackage{comment}
\usepackage[]{units}
\usetikzlibrary{shapes}
\usetikzlibrary{positioning}

\usepackage{silence}
\usepackage[colorlinks=true,backref=false, linktocpage=true,
citecolor=blue,urlcolor=blue,linkcolor=blue,pdfpagemode=UseOutlines]{hyperref}
\hypersetup{%
  bookmarksnumbered=true,
  pdftitle = {Closing the knowledge gap in semileptonic B→Xclν decays},
  pdfsubject = {Semileptonic gap},
  pdfauthor = {Florian Herren, Raynette van Tonder},
  pdfkeywords = {Semileptonic decays, Flavor physics}
}

\allowdisplaybreaks

\begin{document}
\preprint{ZU-TH 17/26}
\title{Closing the knowledge gap in semileptonic \texorpdfstring{$B\rightarrow X_c\ell\nu$}{B→Xclν} decays}

\author{Florian Herren}
\affiliation{Physics Institute, Universität Zürich, Winterthurerstrasse 190, CH-8057 Zürich, Switzerland}

\author{Raynette van Tonder}
\affiliation{Institut für Experimentelle Teilchenphysik, Karlsruhe Institute of Technology (KIT), D-76131 Karlsruhe, Germany}

\date{\today}

\begin{abstract}
In this work we summarize the current status of measured exclusive semileptonic branching fractions containing charm mesons. We use the available experimental data to evaluate the difference between the sum of exclusive measurements and the inclusive determination. By including experimental results of branching fractions relative to semi-inclusive $B\rightarrow D X\ell\nu$ decays, we demonstrate that the unmeasured components of the total branching fraction are dominated by final states devoid of $D$ mesons, hinting towards sizeable contributions from baryonic final states and $D_s$ mesons. Based on the obtained fractions, we discuss candidates that could potentially close the remaining difference and propose searches for promising final states. Furthermore, we provide simplified models for S-wave $B\rightarrow D\eta\ell\nu$ and $B\rightarrow D_s K\ell\nu$ decays that contribute marginally to the unmeasured components of the total inclusive rate.
\end{abstract}

\maketitle

\section{Introduction}
Precise determinations of the absolute value of the Cabibbo-Kobayashi-Maskawa (CKM) matrix elements provide a potent test of the Standard Model (SM) by overconstraining the CKM unitarity triangle~\cite{Charles:2004jd,HeavyFlavorAveragingGroupHFLAV:2024ctg,UTfit:2022hsi}. A well-established strategy to determine $|V_{ub}|$ and $|V_{cb}|$ is through measurements of semileptonic $B$ meson decays with $b \rightarrow u\ell \nu$ and $b \rightarrow c\ell \nu$ transitions. Determinations of $|V_{ub}|$ and $|V_{cb}|$ are extracted by employing two complementary approaches: the exclusive approach focuses on the reconstruction of a specific decay mode, while the inclusive approach aims to measure the sum of all possible final states entailing the same quark-level transition. Current world averages of $|V_{ub}|$ and $|V_{cb}|$ from exclusive and inclusive determinations exhibit disagreements of approximately three standard deviations between both techniques, posing a longstanding puzzle that remains to be resolved.

Determinations of both inclusive and exclusive $|V_{ub}|$ are challenging due to the overwhelming CKM-favored $B \rightarrow X_{c} \ell \nu$ background, which exhibits a similar experimental signature and is $\mathcal{O}(100)$ times more abundant than $B \rightarrow X_{u} \ell \nu$ decays. Methods to separate and suppress background are further complicated by the limited experimental and theoretical knowledge of the orbitally excited $D^{**}$ states, whose branching fractions exhibit uncertainties of approximately 20\%~\cite{ParticleDataGroup:2024cfk}, as well as unmeasured non-resonant $B \rightarrow X_{c} \ell \nu$ decays. The existence of these unmeasured decays results in a non-zero difference between the sum of all known exclusive modes and the total inclusive $B \rightarrow X_{c} \ell \nu$ rate, a problem commonly referred to as the “semileptonic gap”.

In recent measurements performed by the Belle and Belle II Collaborations, the remaining “gap” is generally treated in simulation by assuming a composition of equal parts of $B \rightarrow D^{(*)} \eta \ell \nu $ decays, with the final state usually produced via an intermediate broad S-wave resonance to provide more accurate kinematic description. Due to the lack of experimental evidence for $B \rightarrow D^{(*)} \eta \ell \nu $ decays, a 100\% uncertainty is assumed for the corresponding branching fractions. For these reasons, the $X_{c} \ell \nu$ modeling uncertainty is hard to quantify and becomes dominant for studies of inclusive $B \rightarrow X_{c/u} \ell \nu$ decays~\cite{Belle:2021idw,Belle:2021eni,Belle-II:2025pye}. Therefore, improved knowledge of the largely unmeasured non-resonant processes would not only be essential to provide precise experimental measurements for future extractions of inclusive $|V_{cb}|$, but also allow for improved background suppression for measurements of $|V_{ub}|$. 

Measurements carried out by the LHCb Collaboration assume that the remaining non-resonant contributions consist entirely of heavy excited charm states $D^{**}_{\textrm{heavy}}$ decaying as $D^{**}_{\textrm{heavy}} \rightarrow D^{(*)}\pi\pi$. These backgrounds are modeled using simulated samples containing a variety of possible final states, while the composition and decay properties of this background component are constrained using data-driven control regions~\cite{LHCb:2023zxo}. Despite different treatments of the non-resonant background components, the limited knowledge of $B\rightarrow D^{\ast\ast}\ell/\tau\nu$ branching fractions and form factors are large systematic uncertainties in tests of lepton flavor universality such as $R(D^{(\ast)})$ at both experiments~\cite{LHCb:2023uiv,LHCb:2023zxo,Belle-II:2023aih,Belle-II:2024ami,Belle-II:2025yjp}. Furthermore, measurements of the relative $B^0_s/B^0$ production fraction in semi-inclusive semileptonic decays at LHCb rely on the same assumption \cite{LHCb:2011leg,LHCb:2019fns}, potentially biasing these measurements \cite{Bolognani:2026yxj}.

While both experiments employ different approaches to deal with the remaining unmeasured contributions comprising the total inclusive width, the underlying assumptions regarding these non-resonant decays have been shown to be inadequate. As reported in Ref.~\cite{Gustafson:2023lrz}, the expected branching fraction of $B\rightarrow D\eta\ell\nu$ is two orders of magnitude too small to constitute a sizeable portion of the semileptonic "gap". Since heavy-quark spin symmetry relates the coupled-channel S-wave scattering matrix to the $J^P = 1^+$ S-wave scattering matrix \cite{Du:2017zvv}, the same conclusion holds for the $B \rightarrow (D^\ast \eta)_S \ell \nu$ channel. Consequently, the approach utilized by Belle and Belle II in recent measurements to fill the semileptonic gap in terms of $B\rightarrow D^{(\ast)} \eta \ell \nu$ decays is ruled out. Furthermore, the assumption that the semileptonic gap comprises entirely of heavy excited charm states decaying as $D^{**}_{\textrm{heavy}} \rightarrow D^{(*)}\pi\pi$ would require an unexpectedly large branching fraction of approximately 1\%~\cite{Bernlochner:2012bc}, which would be difficult to understand theoretically~\cite{Becirevic:2012te}. In addition, both Belle and BaBar reported measurements of semileptonic $B \rightarrow D^{(*)} \pi \pi \ell \nu$ final states~\cite{Belle:2022yzd,BaBar:2015zkb} with the resulting branching fractions accounting for only half the difference between the measured inclusive semileptonic branching fraction and the corresponding sum of previously measured exclusive branching fractions.

To address these limitations and improve the precision of future measurements of $|V_{ub}|$ and $|V_{cb}|$, as well as tests of lepton flavor universality, we evaluate the size of the semileptonic gap using existing experimental measurements of exclusive branching fractions in Section~\ref{sec::gapsize}. We simultaneously determine the branching fraction of semi-inclusive $B\rightarrow D X \ell\nu$ decays together with the composition of the semileptonic gap in terms of different charm hadrons. Next, in Section~\ref{sec::swave} we provide updated branching fraction predictions and improved descriptions of S-wave $B\rightarrow D\pi/D\eta/D_s K\ell\nu$ decays that contribute marginally to the unmeasured components of the total inclusive rate. Section~\ref{sec::candidates} presents a non-exhaustive list of candidates that could potentially contribute to the semileptonic gap together with additional motivation for direct searches of these modes. In Section~\ref{sec::Recommendations} we provide recommendations for the treatment of the gap in Monte Carlo simulations in future experimental analyses and highlight several modes of particular interest. We provide an outlook in Section~\ref{sec::Outlook}.

\section{Determining the size of the semileptonic gap}\label{sec::gapsize}
In the following, we determine the size of the semileptonic gap separately for charged and neutral $B$ mesons. We refrain from utilizing isospin averages, as complications related to effects of the $D^\ast$ tail~\cite{LeYaouanc:2018zyo,LeYaouanc:2021xcq,Gustafson:2023lrz,Du:2025beb} can not be fully controlled at present.
Similarly, we do not rely on any experimental results of specific heavy excited charm states, since all of the existing measurements of these branching fractions rely on model-dependent fits to the hadronic invariant mass spectra.

Throughout this work, $D^{\ast\ast}$ refers to any hadronic final state, resonant or non-resonant, including $D$ mesons. Furthermore, we follow the resonance naming scheme employed within the Review of Particle Physics~\cite{ParticleDataGroup:2024cfk}: $D_J^{(\ast)}(M)$.
Here, $J$ denotes the angular momentum of the resonance and $M$ the nominal mass. An asterisk is added for resonances with natural parity, i.e. for states with $(-1)^J = P$, such as the $2^+$ $D_2^\ast(2460)$ or the $1^-$ $D_1^\ast(2600)$.

\subsection{Experimental inputs}
We take the HFLAV averages~\cite{HeavyFlavorAveragingGroupHFLAV:2024ctg} for the branching ratios of the dominant exclusive modes $B\rightarrow D^{(\ast)}\ell\nu$:
\begin{align}
    \mathcal{B}(B^+\rightarrow \bar{D}^0\ell^+\nu_\ell) &= (2.12 \pm 0.02 \pm 0.06)\%~,\nonumber\\
    \mathcal{B}(B^0\rightarrow D^-\ell^+\nu_\ell) &= (2.21 \pm 0.02 \pm 0.06)\%~,\nonumber\\
    \mathcal{B}(B^+\rightarrow \bar{D}^{\ast 0}\ell^+\nu_\ell) &= (5.53 \pm 0.07 \pm 0.21)\%~,\nonumber\\
    \mathcal{B}(B^0\rightarrow D^{\ast -}\ell^+\nu_\ell) &= (4.90 \pm 0.01 \pm 0.12)\%~.
    \label{eq::DDastBFs}
\end{align}
Modes involving one or two additional pions have been measured by the Belle~\cite{Belle:2022yzd} and BaBar~\cite{BaBar:2007ddh,BaBar:2015zkb} experiments. The Belle measurement reports branching fraction ratios with respect to the  $B\rightarrow D^\ast\ell\nu$ normalization mode, from which the resulting branching fractions are obtained. Similarly, BaBar uses $B\rightarrow X\ell\nu$ as a normalization mode for the single pion modes and $B\rightarrow D^{(\ast)}\ell\nu$ for the di-pion modes.
In our analysis, we directly use the measured ratios to retain information on the correlations between different exclusive modes. We neglect the BaBar measurement of the single-pion modes, since the results have limited precision compared to the Belle measurement and carries little weight in an average. In summary, we use the following experimental results~\cite{Belle:2022yzd}:
\begin{align}
    \frac{\mathcal{B}(B^+\rightarrow D^-\pi^+\ell^+\nu_\ell)}{\mathcal{B}(B^+\rightarrow \bar{D}^{\ast 0}\ell^+\nu_\ell)} &= (6.78 \pm 0.24 \pm 0.15)\%~,\nonumber\\
    \frac{\mathcal{B}(B^0\rightarrow \bar{D}^0\pi^-\ell^+\nu_\ell)}{\mathcal{B}(B^0\rightarrow D^{\ast -}\ell^+\nu_\ell)} &= (7.24 \pm 0.36 \pm 0.12)\%~,\nonumber\\
    \frac{\mathcal{B}(B^+\rightarrow D^{\ast -}\pi^+\ell^+\nu_\ell)}{\mathcal{B}(B^+\rightarrow \bar{D}^{\ast 0}\ell^+\nu_\ell)} &= (9.50 \pm 0.33 \pm 0.27)\%~,\nonumber\\
    \frac{\mathcal{B}(B^0\rightarrow \bar{D}^{\ast 0}\pi^-\ell^+\nu_\ell)}{\mathcal{B}(B^0\rightarrow D^{\ast -}\ell^+\nu_\ell)} &= (11.10 \pm 0.48 \pm 0.20)\%~,\nonumber\\
    \frac{\mathcal{B}(B^+\rightarrow \bar{D}^0\pi^+\pi^-\ell^+\nu_\ell)}{\mathcal{B}(B^+\rightarrow \bar{D}^{\ast 0}\ell^+\nu_\ell)} &= (3.10 \pm 0.26 \pm 0.21)\%~,\nonumber\\
    \frac{\mathcal{B}(B^0\rightarrow D^-\pi^+\pi^-\ell^+\nu_\ell)}{\mathcal{B}(B^0\rightarrow D^{\ast -}\ell^+\nu_\ell)} &= (2.91 \pm 0.37 \pm 0.25)\%~,\nonumber\\
    \frac{\mathcal{B}(B^+\rightarrow D^{\ast 0}\pi^+\pi^-\ell^+\nu_\ell)}{\mathcal{B}(B^+\rightarrow \bar{D}^{\ast 0}\ell^+\nu_\ell)} &= (1.25 \pm 0.27 \pm 0.15)\%~,\nonumber\\
    \frac{\mathcal{B}(B^0\rightarrow \bar{D}^{\ast -}\pi^+\pi^-\ell^+\nu_\ell)}{\mathcal{B}(B^0\rightarrow D^{\ast -}\ell^+\nu_\ell)} &= (1.03 \pm 0.43 \pm 0.18)\%
\end{align}
and~\cite{BaBar:2015zkb}
\begin{align}
    \frac{\mathcal{B}(B^+\rightarrow \bar{D}^0\pi^+\pi^-\ell^+\nu_\ell)}{\mathcal{B}(B^+\rightarrow \bar{D}^{0}\ell^+\nu_\ell)} &= (7.1 \pm 1.3 \pm 0.8)\%~,\nonumber\\
    \frac{\mathcal{B}(B^0\rightarrow D^-\pi^+\pi^-\ell^+\nu_\ell)}{\mathcal{B}(B^0\rightarrow D^{-}\ell^+\nu_\ell)} &= (5.8 \pm 1.8 \pm 1.2)\%~,\nonumber\\
    \frac{\mathcal{B}(B^+\rightarrow D^{\ast 0}\pi^+\pi^-\ell^+\nu_\ell)}{\mathcal{B}(B^+\rightarrow \bar{D}^{\ast 0}\ell^+\nu_\ell)} &= (1.4 \pm 0.7 \pm 0.4)\%~,\nonumber\\
    \frac{\mathcal{B}(B^0\rightarrow \bar{D}^{\ast -}\pi^+\pi^-\ell^+\nu_\ell)}{\mathcal{B}(B^0\rightarrow D^{\ast -}\ell^+\nu_\ell)} &= (2.8 \pm 0.8 \pm 0.6)\%~.
\end{align}
We denote the above-mentioned ratios as
\begin{align}
    r^{(\ast)}_\pi &= \frac{\mathcal{B}(B\rightarrow D^{(\ast)}\pi^\pm\ell\nu_\ell)}{\mathcal{B}(B\rightarrow D^{\ast}\ell\nu_\ell)}~,\nonumber\\
    r^{(\ast)}_{\pi\pi} &= \frac{\mathcal{B}(B\rightarrow D^{(\ast)}\pi^+\pi^-\ell\nu_\ell)}{\mathcal{B}(B\rightarrow D^{\ast}\ell\nu_\ell)}~.
\end{align}

Limited studies of semileptonic $B$ decays into charm hadrons besides $D$ mesons are currently available. However, BaBar~\cite{BaBar:2010ner} and Belle~\cite{Belle:2012ccr} observed $B^+\rightarrow D_s^{(\ast)-}K^+\ell^+\nu_\ell$ decays in the combination of $D_s$ and $D_s^{(\ast)}$ final states. In addition, Belle found evidence for the $D_s$ mode with a significance of 3.4$\sigma$ and consequently provided a measurement of the branching fraction. We averaging the results and obtain
\begin{align}
    \mathcal{B}(B^+\rightarrow D^-_s K^+ \ell^+\nu_\ell) &= (3.05 \pm 1.15)\times 10^{-4}~,\nonumber\\
    \mathcal{B}(B^+\rightarrow D^{\ast-}_s K^+ \ell^+\nu_\ell) &= (3.01\pm 1.24)\times 10^{-4} ~,
\end{align}
with a correlation coefficient of $\rho = -0.69$. Their combination leads to
\begin{align}
    \mathcal{B}(B^+\rightarrow D^{(\ast)-}_s K^+ \ell^+\nu_\ell) &= (6.06 \pm 0.95)\times 10^{-4}~.
\end{align}
Assuming isospin symmetry for the corresponding decay width, we convert the branching ratio to $B^0$ decays by making use of the lifetime ratio~\cite{HeavyFlavorAveragingGroupHFLAV:2024ctg}:
\begin{align}
\tau(B^+)/\tau(B^0) &= 1.076 \pm 0.004~.
\end{align}
We obtain
\begin{align}
    \mathcal{B}(B^0\rightarrow D^{(\ast)-}_s K^0 \ell^+\nu_\ell) &= (5.63 \pm 0.88)\times 10^{-4}~.
\end{align}

BaBar~\cite{BaBar:2007xlq} and LHCb~\cite{LHCb:2018azb} report branching fractions relative to the semi-inclusive $B\rightarrow D X \ell\nu$ rate, denoted as $f_{D/D^\ast/D^{\ast\ast}} = \mathcal{B}(B\rightarrow D/D^\ast/D^{\ast\ast}\ell\nu)/\mathcal{B}(B\rightarrow D X \ell\nu)$. Since the LHCb measurements are compatible, but less precise than the results report by BaBar, we only include the latter in our estimations. We use \cite{BaBar:2007xlq}
\begin{align}
    f_D^+ = (22.7 \pm 1.4 \pm 1.6) \%~,\nonumber\\
    f_D^0 = (21.5 \pm 1.6 \pm 1.3) \%~,\nonumber\\
    f_{D^\ast}^+ = (58.2 \pm 1.8 \pm 3.0) \%~,\nonumber\\
    f_{D^\ast}^0 = (53.7 \pm 3.1 \pm 3.6) \%~,\nonumber\\
    f_{D^{\ast\ast}}^+ = (19.1 \pm 1.3 \pm 1.9) \%~,\nonumber\\
    f_{D^{\ast\ast}}^0 = (24.8 \pm 3.2 \pm 3.0) \%~,\label{eq::babarfracs}
\end{align}
where the superscript denotes the charge of the $B$ meson. The ratio $f_D/f_D^\ast$ is compatible within uncertainties with the ratios of branching fractions given in Eq.~\eqref{eq::DDastBFs} for both charge modes.

Lastly, the final ingredient is the inclusive $B\rightarrow X_c\ell\nu$ branching fraction, which we take from the global fit to kinematic moments of Ref.~\cite{Finauri:2023kte}:
\begin{align}
    \mathcal{B}(B\rightarrow X_c\ell\nu) &= (10.63\pm 0.15) \%~.
\end{align}

\subsection{Fit setup}
To determine the $B\rightarrow DX\ell\nu$ branching fraction as well as the size of modes yet unaccounted for, such as modes with three or more pions, $\eta^{(\prime)}$ mesons, kaon pairs or photons, we need to extrapolate the single-pion and di-pion modes to account for neutral pions.

Assuming isospin symmetry, the total $B\rightarrow D^{(\ast)}\pi\ell\nu$ branching fractions are directly proportional to the case of charged pions:
\begin{align}
    \mathcal{B}(B\rightarrow D^{(\ast)}\pi\ell\nu) = \frac{3}{2}\mathcal{B}(B\rightarrow D^{(\ast)}\pi^\pm\ell\nu)~.
\end{align}
Small corrections to this ratio can arise near the $D^{(\ast)}\pi$ threshold, in particular for neutral $D^{(\ast)}\pi$ systems, due to the mass differences and the nearby $D^\ast$ resonance. However, these effects should be covered by current uncertainties.

The di-pion modes proceed through several subprocesses, all of which exhibiting non-trvial interference patterns. We consider three scenarios to estimate the extrapolation factor, following a similar prescription as Refs.~\cite{Lueck2013Determination,BaBar:2015zkb,Bernlochner:2016bci,Rudolph:2018rzl}:
\begin{enumerate}
    \item Sequential decays of the form $D^{\ast\ast}\rightarrow D^{\prime\ast\ast}(\rightarrow D^{(\ast)}\pi)\pi$. This case also includes a component from the $D^\ast$ tail, a fraction of which will contribute to the $D\pi\pi$ channel due to the $D^\ast\pi$ modes. Furthermore, Belle reported a preference for a dominant S-wave $D\pi$ component in $D_1(2420)\rightarrow D\pi\pi$ decays~\cite{Belle:2004bvv}. In this case, the measured $D^{(\ast)}\pi^+\pi^-$ modes account for $4/9$ of the total $D^{(\ast)}\pi\pi$ contribution.\footnote{This relies on the assumption that no large intermediate contributions from isospin-$3/2$ charm resonances exist. To date, no such states have been observed.}
    \item $D^{\ast\ast}\rightarrow D^{(*)}(\pi\pi)_{I=0}$ decays, which implies that the measured modes account for $2/3$ of the total contribution. This mode could contribute significantly to $D_1^\ast(2600)$ decays through the S-wave.
    \item $D^{\ast\ast}\rightarrow D^{(*)}(\pi\pi)_{I=1}$ decays, which could be the dominant mode for invariant masses above $2.8$ GeV, since the $\rho$ peak would be fully in the phase space. In this case, the observed modes account for $1/3$ of the total amount.
\end{enumerate}
Treating all three scenarios as equal possibilities, we determine
\begin{align}
    f^{(\ast)}_{\pi\pi} \equiv \frac{\mathcal{B}(B\rightarrow D^{(\ast)}\pi\pi\ell\nu)}{\mathcal{B}(B\rightarrow D^{(\ast)}\pi^+\pi^-\ell\nu)} = \frac{9}{4}\pm \frac{3}{4}~.
\end{align}
We treat the $D$ and $D^\ast$ extrapolation uncorrelated.

The observed $B\rightarrow D X\ell\nu$ branching fraction is defined by
\begin{align}
    &\mathcal{B}(B\rightarrow D X\ell\nu)\Big|_\text{obs} = \mathcal{B}(B\rightarrow D\ell\nu) \nonumber\\&+ \mathcal{B}(B\rightarrow D^\ast\ell\nu)\left(1 +\frac{3}{2}r_\pi+ \frac{3}{2}r^\ast_\pi+f_{\pi\pi}r_{\pi\pi}+f^\ast_{\pi\pi}r^\ast_{\pi\pi}\right)\,,
\end{align}
where we obtain the branching fractions of decays involving additional pions from the measured ratios.

The total $B\rightarrow D X\ell\nu$ branching fraction is given by
\begin{align}
    \mathcal{B}(B\rightarrow D X\ell\nu) &= \mathcal{B}(B\rightarrow D X\ell\nu)\Big|_\text{obs}\nonumber\\ &+ \mathcal{B}(B\rightarrow D X\ell\nu)\Big|_\text{gap}~,
\end{align}
where $\mathcal{B}(B\rightarrow D X\ell\nu)\Big|_\text{gap}$ will be constrained through the relative fractions in Eq.~\eqref{eq::babarfracs}. 

In addition to final states containing $D$ mesons, $b\rightarrow c$ transitions also result in final states that include $D_s$ mesons or $\Lambda_c$ baryons. We introduce an additional component that can be further split into a second gap component, as well as the observed $B\rightarrow D_s^{(\ast)} K \ell\nu$ branching fraction:
\begin{align}
    \mathcal{B}(B\rightarrow D_s/\Lambda_c\,X \ell\nu) &= \mathcal{B}(B\rightarrow D_s/\Lambda_c\,X \ell\nu)\Big|_\text{gap}\nonumber\\ &+ \mathcal{B}(B\rightarrow D_s^{(\ast)} K \ell\nu)~.
\end{align}
Note, that decays into $\Xi_c^{(\prime)}$ and $\Omega_c$ baryons are also possible, however due to their higher mass they will be suppressed compared to decays into $\Lambda_c$ baryons.

Finally, the inclusive branching fraction is the sum of the two components:
\begin{align}
    \mathcal{B}(B\rightarrow X_c\ell\nu) = \mathcal{B}(B\rightarrow D X\ell\nu) + \mathcal{B}(B\rightarrow D_s/\Lambda_c\,X \ell\nu)~.
\end{align}

We now study the available data in a Bayesian analysis. The posterior probability distribution function (PDF) $\mathcal{P}(\vec{x}|\text{data})$ is a function of the available data and our fit parameters $\vec{x}$. These parameters include: the branching fractions $\mathcal{B}(B\rightarrow D^{(\ast)}\ell\nu_\ell)$, the ratios $r_{\pi(\pi)}^{(\ast)}$, the isospin factors $f^{(\ast)}_{\pi\pi}$, as well as the two parameters of interest $\mathcal{B}(B\rightarrow D X\ell\nu)\Big|_\text{gap}$ and $\mathcal{B}(B\rightarrow D_s/\Lambda_c\,X \ell\nu)$. The posterior is defined by
\begin{align}
    \mathcal{P}(\vec{x}|\text{data}) = \frac{\mathcal{P}(\text{data}|\vec{x})\mathcal{P}_0(\vec{x})}{Z(\text{data})}~,
\end{align}
where the normalization $Z(\text{data})$ is called the evidence, $\mathcal{P}(\text{data}|\vec{x})$ is the likelihood and $\mathcal{P}_0(\vec{x})$ is the prior.
With the exception of the priors for the isospin factors $f^{(\ast)}_{\pi\pi}$ and the gap branching fractions, which we assume to be uniform, all priors are assumed to be gaussian. We study the posterior through the nested sampling algorithm \cite{Higson:2018cwj} implemented in \verb|dynesty| \cite{Speagle:2019ivv}.

\subsection{The size of the gap}
We consider two scenarios: in the first, we determine values for $\mathcal{B}(B\rightarrow D X \ell\nu)$ and $\mathcal{B}(B\rightarrow D_s/\Lambda_c\,X \ell\nu)$ independently for $B^+$ and $B^0$ decays, while in the second we constrain both to common values assuming isospin symmetry. For both scenarios, the posteriors for $\mathcal{B}(B\rightarrow D^{(\ast)}\ell\nu)$ and $r_{\pi(\pi)}^{(\ast)}$ reproduce the priors. The isospin factors $f^{(\ast)}_{\pi\pi}$ show a slight preference for smaller (larger) values for $B^+$ ($B^0$) decays in the second scenario.

In the first scenario, the size of the gap components is determined to be
\begin{align}
    \mathcal{B}(B^+\rightarrow D_s/\Lambda_c\,X \ell^+\nu_\ell)\Big|_\mathrm{gap} &= (1.29\pm 0.37)\%~,\nonumber\\
    \mathcal{B}(B^+\rightarrow D X \ell^+\nu_\ell)\Big|_\mathrm{gap} &= (0.44\pm 0.23)\%~,
\end{align}
for the $B^+$ decays and
\begin{align}
    \mathcal{B}(B^0\rightarrow D_s/\Lambda_c\,X \ell^+\nu_\ell)\Big|_\mathrm{gap} &= (0.83\pm 0.42)\%~,\nonumber\\
    \mathcal{B}(B^0\rightarrow D X \ell^+\nu_\ell)\Big|_\mathrm{gap} &= (1.01\pm 0.38)\%~,
\end{align}
for $B^0$ decays. The values of $\mathcal{B}(B\rightarrow D_s/\Lambda_c\,X \ell\nu)|_\mathrm{gap}$ are compatible with isospin symmetry within uncertainties.
Combining the two separate gap components, we obtain
\begin{align}
    \mathcal{B}(B^+\rightarrow X_c \ell^+\nu_\ell)\Big|_\mathrm{gap} &= (1.78 \pm 0.33)\%~,\nonumber\\
    \mathcal{B}(B^0\rightarrow X_c \ell^+\nu_\ell)\Big|_\mathrm{gap} &= (1.81 \pm 0.24)\%~.
\end{align}
In addition, the resulting $B\rightarrow D X\ell\nu$ branching fractions are
\begin{align}
    \mathcal{B}(B^+\rightarrow D X \ell^+\nu_\ell) &= (9.61 \pm 0.35)\%~,\nonumber\\
    \mathcal{B}(B^0\rightarrow D X \ell^+\nu_\ell) &= (9.39 \pm 0.39)\%~.\label{eq::dxseparate}
\end{align}

\begin{figure}[ht]
    \centering
    \includegraphics[width=0.48\textwidth]{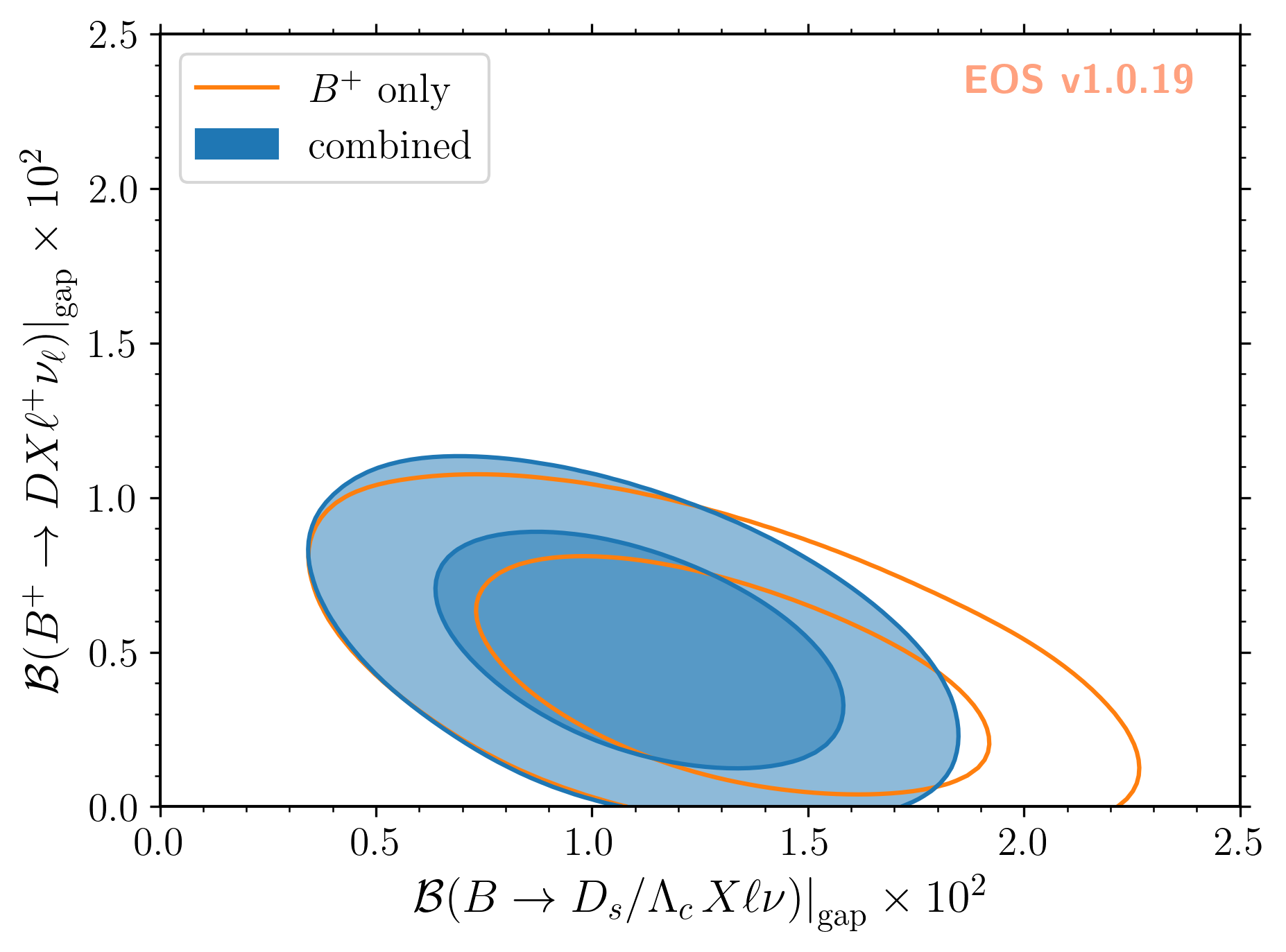}
    \includegraphics[width=0.48\textwidth]{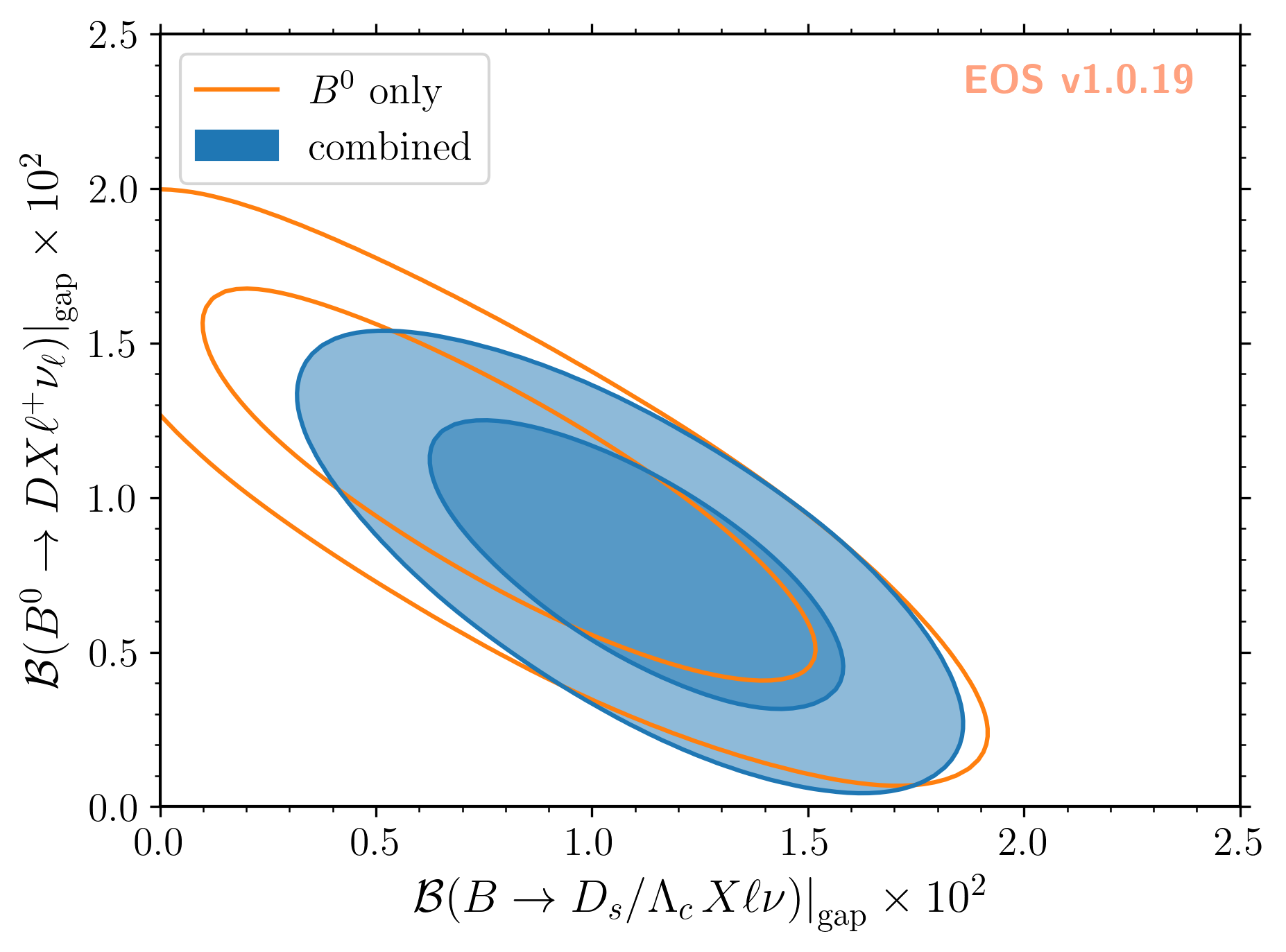}
    \caption{\label{fig:gapbfs}
        The size of the separate semileptonic gap components for $B^{+}$ decays (top) and $B^{0}$ decays (bottom) for the two considered fit scenarios in which the components are determined independently for $B^{+}$ and $B^{0}$ decays (orange) and assuming isospin symmetry (blue). The inner and outer ellipses correspond to the 1 and 2 $\sigma$ contours, respectively. The figures were generated with the plotting framework of EOS v1.0.19~\cite{EOSAuthors:2021xpv,the_eos_authors_collaboration_2025_17792609}.
    }
\end{figure}

In the second scenario, the three gap components are more precise but compatible with the results of the first scenario:
\begin{align}
    \mathcal{B}(B\rightarrow D_s/\Lambda_c\,X \ell\nu)\Big|_\mathrm{gap} &= (1.10\pm 0.29)\%~,\ \nonumber\\
    \mathcal{B}(B^+\rightarrow D X \ell^+\nu_\ell)\Big|_\mathrm{gap} &= (0.52\pm 0.23)\%~,\ \nonumber\\
    \mathcal{B}(B^0\rightarrow D X \ell^+\nu_\ell)\Big|_\mathrm{gap} &= (0.79\pm 0.28)\%~.
    \label{eq::fitresults}
\end{align}
The total $D X \ell\nu$ contribution is very close to the average of the separate $B^+$ and $B^0$ results in Eq.~\eqref{eq::dxseparate}:
\begin{align}
\mathcal{B}(B\rightarrow DX \ell\nu)\phantom{\Big|_\mathrm{gap}} &= (9.47\pm 0.27)\%~.
\end{align}
This value is in perfect agreement with the findings of Ref.~\cite{Bolognani:2026yxj}.
Near the best-fit-point, the uncertainties are approximately Gaussian and the correlation matrix of the four quantities is given by
\begin{align}
    C =
    \begin{pmatrix}
     1   & -0.49 & -0.72 & -0.93~ \\
    -0.49 &  1    &  0.40 &  0.52 \\
    -0.72 &  0.40 &  1    &  0.77 \\
    -0.93 &  0.52 &  0.77 &  1
    \end{pmatrix}~.
\end{align}
While the central values of the different gap components shift between the two scenarios, the results of both fits are in good agreement with each other as demonstrated in Fig.~\ref{fig:gapbfs}.

Our results yield a non-vanishing branching ratio into $D_s$ or baryonic final states at the $99\%$ confidence level, indicating a significant $X_c$ component that does not result in final states involving $D$ mesons. The resulting branching fraction can be compared to upper limits on $B\rightarrow D_s X \ell\nu$ and $B\rightarrow \Lambda_c\,X\ell\nu$ decays.
For $D_s$ final states, ARGUS obtained an upper limit of~\cite{ARGUS:1993hhz}
\begin{align}
    \mathcal{B}(B\rightarrow D_s X \ell\nu) < 1.2\% ,\
\end{align}
at the $90\%$ confidence level, whereas BaBar sets a limit on the ratio~\cite{BaBar:2011ijz}
\begin{align}
    \frac{\mathcal{B}(B\rightarrow \Lambda_c^-X\ell^+\nu_\ell)}{\mathcal{B}(B\rightarrow\Lambda_c^-X)} < 0.035 ,\
\end{align}
at the $90\%$ confidence level.
The two limits are rescaled in the Review of Particle Physics to~\cite{ParticleDataGroup:2024cfk}
\begin{align}
    \mathcal{B}(B\rightarrow D_s\,X \ell\nu) &< 0.7\%\nonumber~,\\
    \mathcal{B}(B\rightarrow \Lambda_c\,X\ell\nu) &< 0.09 \%~.\label{eq::lcX}
\end{align}
Combining both limits results in an approximate upper limit that is in slight tension with our fit results:
\begin{align}
    \mathcal{B}(B\rightarrow D_s/\Lambda_c\,X \ell\nu) &< 0.8\%\nonumber~.
\end{align}

\section{S-wave \texorpdfstring{$B\rightarrow D\eta\ell\nu$}{B→Detalν} and \texorpdfstring{$B\rightarrow D_s K\ell\nu$}{B→DsKlν} decays}\label{sec::swave}
The size of the S-wave $D\eta$ and $D_s K$ final states can be constrained in a data-driven manner through measurements of the $D\pi$ invariant mass spectrum in $B\rightarrow D\pi\ell\nu$ decays and a coupled-channel description of $D\pi-D\eta-D_sK$ S-wave scattering, as first shown in Ref.~\cite{Gustafson:2023lrz}.

\subsection{Lineshapes and form factors}
The matrix element relevant for charged-current semileptonic $B$ decays into a $J^P = 0^+$ final state $X$ in the limit of massless leptons is given by~\cite{Gustafson:2023lrz,Herren:2025cwv,Du:2025beb}
\begin{align}
    \left\langle X | A^\mu | B \right\rangle = \left[(p_B +p_X)^\mu + \frac{M_B^2 - M_X^2}{q^2}q^\mu\right]f_X(q^2,M_X^2)~,
\end{align}
with $q^\mu = p_B^\mu-p_X^\mu$. Combining the three channels $X = D\pi, D\eta, D_sK$, their $M_X$-dependence is determined through \cite{Du:2025beb}
\renewcommand\arraystretch{1.2}
\begin{align}
    \begin{pmatrix}
    f_{D\pi}(q^2,M_X^2) \\ f_{D\eta}(q^2,M_X^2) \\ f_{D_s K}(q^2,M_X^2)
    \end{pmatrix}
    = T(M_X^2)
    \begin{pmatrix}
    \hat{f}_{D\pi}(q^2) \\ \hat{f}_{D\eta}(q^2) \\ \hat{f}_{D_s K}(q^2)
    \end{pmatrix}~,
\end{align}
\renewcommand\arraystretch{1}
where $T$ is the coupled-channel isospin $1/2$ $T$-matrix from Ref.~\cite{Liu:2012zya}.
In the chiral limit, the system can be further simplified~\cite{Du:2025beb}
\begin{align}
    \begin{pmatrix}
    \hat{f}_{D\pi}(q^2) \\ \hat{f}_{D\eta}(q^2) \\ \hat{f}_{D_s K}(q^2)
    \end{pmatrix}
    =\begin{pmatrix}
    \sqrt{3/2} \\ 1/\sqrt{6} \\ 1
    \end{pmatrix} P(q^2)~,
\end{align}
resulting in one effective form factor $P$, which can be parametrized as~\cite{Gustafson:2023lrz}
\begin{align}
    P(q^2) = \frac{1}{B_f(q^2)}\sum_{i=0}^\infty a_i z^i~.\label{eq::swave_ff}
\end{align}
The variable $z$ is given by
\begin{align}
    z = \frac{\sqrt{q^2_+ - q^2} - \sqrt{q^2_+ - q_0^2}}{\sqrt{q^2_+ - q^2} + \sqrt{q^2_+ - q_0^2}}~.
\end{align}
For the S-wave form factor, $q^2_+=(M_{B^\ast} + M_D)^2$ and we choose $q^2_0 = 0$. The Blaschke product $B_f$ incorporates all subthreshold axial-vector $B_c$-meson resonances.

For the P-wave $D^\ast$ tail and the D-wave $D_2^\ast(2460)$ contributions we factorize the $q^2$ and invariant mass dependence of the form factors
\begin{align}
    \mathcal{F}_l(q^2,M_{D\pi}^2)\approx \hat{\mathcal{F}}_l(q^2)h_l(M_{D\pi}^2)~,
\end{align}
where $l = 1$ corresponds to the P-wave and $l = 2$ to the D-wave.
Following Ref.~\cite{Du:2025beb}, we describe the P-wave lineshape through
\begin{align}
    h_1(M_{D\pi}^2) = \frac{g_1 F^{(1)}(M_{D\pi}^2,q_0)}{(M_{D\pi}^2 - M_{D^\ast}^2) + i M_{D^\ast}\Gamma_{D^\ast}(M_{D\pi}^2)}~,
\end{align}
where $g_1$ denotes the $D^\ast-D-\pi$ coupling, $M_{D^\ast}$ the $D^\ast$ mass, $F^{(1)}$ the $L = 1$ Blatt--Weisskopf damping factor~\cite{VonHippel:1972fg}, $q_0 = 0.5~\text{GeV}$ is a fixed momentum scale, and for the neutral $D^\ast$,
\begin{align}
    i\Gamma_{D^\ast}(s) &= i\Gamma_{D^\ast\rightarrow D\gamma} + \frac{1}{M_{D^\ast}}\Bigg[\frac{g^2}{2}\Sigma^{D^0\pi^0}_1(s)\nonumber\\ &+ g^2\Big(\Sigma^{D^+\pi^-}_1(s)-\Sigma^{D^+\pi^-}_1(M^2_{D^\ast})\Big)\Bigg]~.
\end{align}
A similar expression holds for the charged $D^\ast$, however no subtraction is required for either pion channel, since both thresholds lie below the charged $D^\ast$ mass.
The relevant Chew--Mandelstam functions $\Sigma$ are explicitly derived in Ref.~\cite{Du:2025beb}.
We treat the lineshape of the D-wave $D_2^\ast$ in a simplified manner, following Ref.~\cite{Gustafson:2023lrz}:
\begin{align}
    h_2(M_{D\pi}^2) = \frac{F^{(2)}(M_{D\pi}^2,q_0)}{(M_{D\pi}^2 - M_{D_2^\ast}^2) + i M_{D_2^\ast}\Gamma_{D_2^\ast}(M_{D\pi}^2)}~.
\end{align}
The overall normalization is absorbed into the form factors and $\Gamma_{D_2^\ast}(s)$ is taken to be an energy-dependent width, with the assumption that it is saturated by the observed $D\pi$ and $D^\ast\pi$ final states.

\subsection{Fit setup and results}
We perform a least-squares fit employing the package \verb|lsqfit| to the $B\rightarrow D\pi\ell\nu$ invariant mass spectra measured by Belle~\cite{Belle:2022yzd} and use the augmented $\chi^2$ defined in Refs.~\cite{Lepage:2001ym,Hornbostel:2011hu} to judge the quality of the fit.
We consider data only up to $M_{D\pi} = 2.55~\text{GeV}$ to avoid influence from possible additional resonances at higher invariant masses.

The parameters entering the P- and D-wave form factors are determined in fits to the lattice calculation by the Fermilab-MILC collaboration~\cite{FermilabLattice:2021cdg} and kinematic spectra provided by Belle~\cite{Belle:2007uwr}, described in Ref.~\cite{Gustafson:2023lrz}. The choice of input for the $B\rightarrow D^\ast$ form factors has little impact on the results presented in this section, since we only study the $M_{D\pi}$ spectrum.

To obtain the S-wave branching fractions, we truncate the series in Eq.~\eqref{eq::swave_ff} at linear order. Furthermore, we derive uncorrelated shape variations of the $T$-matrix from the eigenvariations of the corresponding parameters, which we include in the fit through nuisance parameters.

\begin{figure}[ht]
    \centering
    \includegraphics[width=0.4825\textwidth]{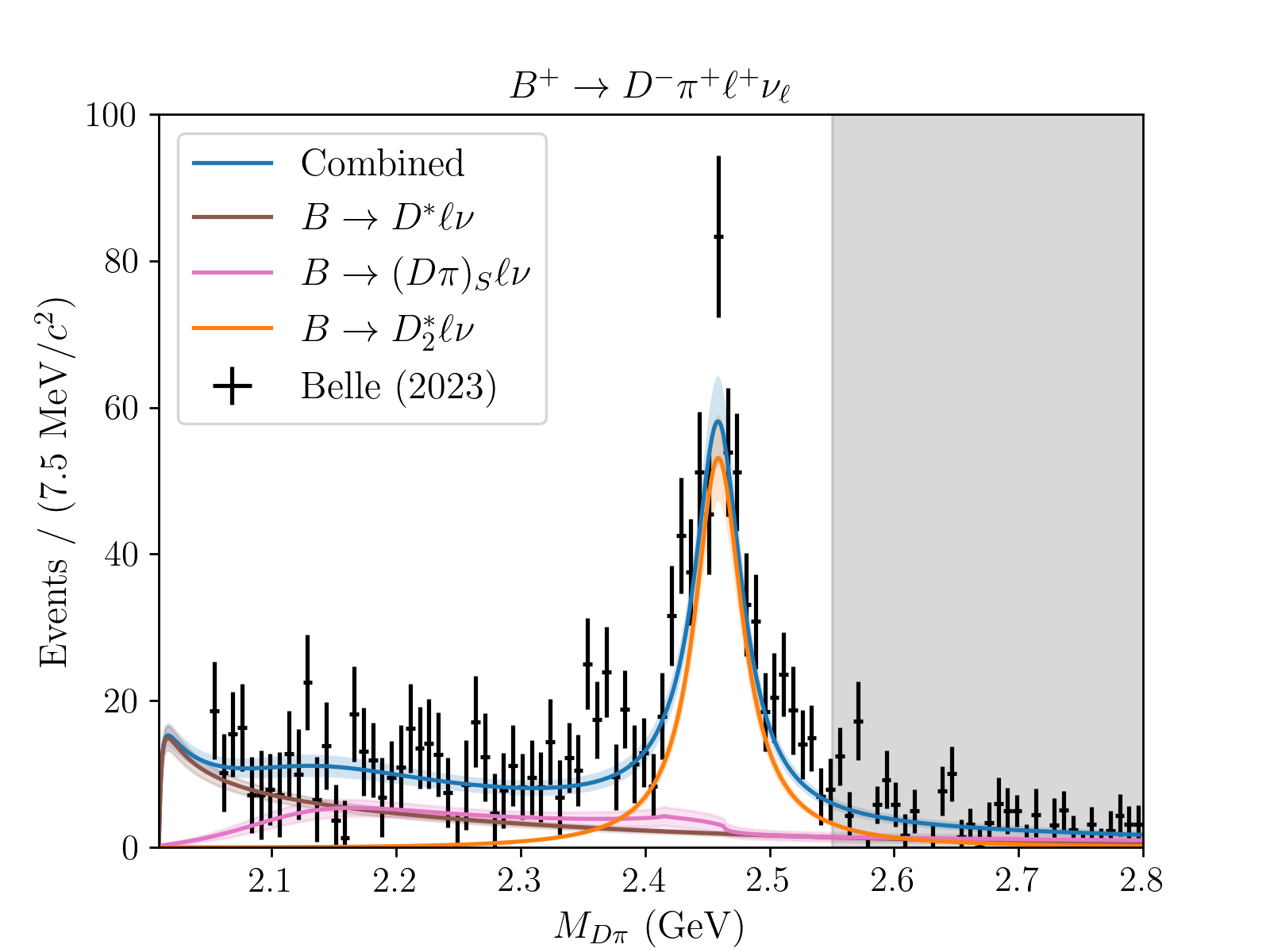}
    \caption{\label{fig:Fit}
        Fit of the measured $M_{D\pi}$-spectrum~\cite{Belle:2022yzd} using the $z$-expansion from Ref.~\cite{Gustafson:2023lrz} to parametrize $D_2^\ast(2460)$ and S-wave form factors using the improved description from Refs.~\cite{Du:2025beb,Herren:2025cwv}. The fit quality is excellent over the entire fit range and even the excluded region beyond $M_{D\pi} \geq 2.55$~GeV with $\chi^2/\text{dof} = 1.0\,(133)$ and $Q = 0.41$. The model predicts a non-vanishing P-wave component in the near-threshold region below $2.05$~GeV due to the subthreshold $D^\ast$.
    }
\end{figure}

Our model provides a good description over the entire fit range and even beyond, as can be seen in Fig.~\ref{fig:Fit}. The fit quality is excellent with $\chi^2/\text{dof} = 1.0\,(133)$ and $Q = 0.41$.

\begin{figure}[ht]
    \centering
    \includegraphics[width=0.4825\textwidth]{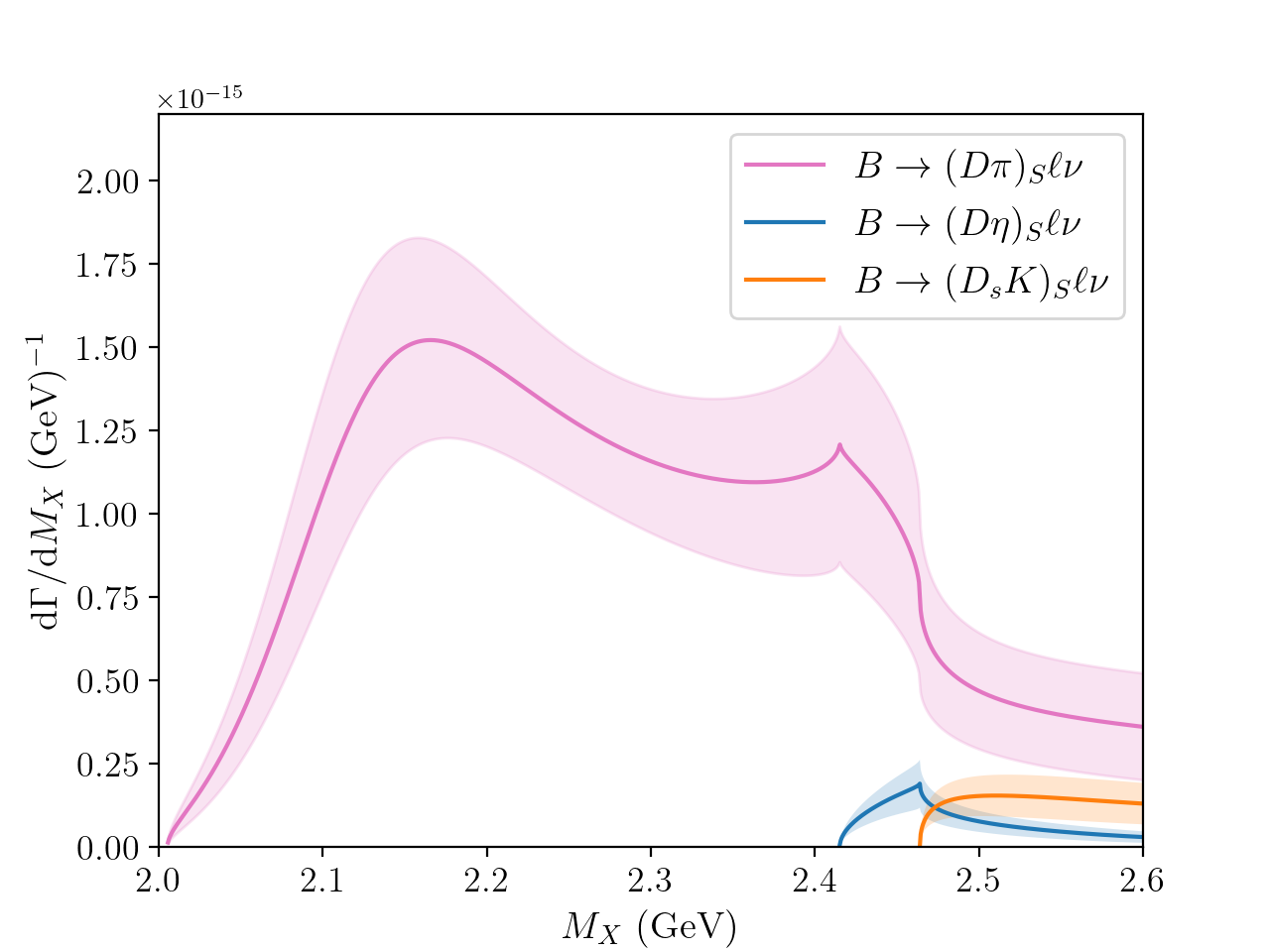}
    \caption{\label{fig:Swaves}
        Predicted lineshapes of the $D\pi$ (pink), $D\eta$ (blue), and $D_sK$ (orange) S-wave contributions from the fit results shown in Fig.~\ref{fig:Fit}.
    }
\end{figure}

The predicted shapes of the S-wave contributions are intricate, as shown in Fig.~\ref{fig:Swaves}. The $D\pi$ lineshape from the $T$-matrix of Ref.~\cite{Liu:2012zya} exhibits two peaks: one near $2.15$~GeV and one near the $D\eta/D_sK$ thresholds at $2.4$~GeV. They originate from two separate poles \cite{Albaladejo:2016lbb,Meissner:2020khl} exhibiting non-trivial interference structures. In addition, the lineshape features two cusps at the $D\eta$ and $D_s K$ thresholds. The $D\eta$ channel rises quickly below the $D_s K$ threshold due to the nearby pole, but falls off steeply after the $D_s K$ channel becomes kinematically allowed. This behavior can be understood directly from the resonance couplings determined in Ref.~\cite{Albaladejo:2016lbb}, since the second pole couples more than twice as strong to the $D_s K$ channel, than the $D \eta$ channel. Lastly, the $D_s K$ channel follows the expected behavior of an S-wave with a subthreshold pole and rises quickly at threshold.
In the remainder of this paper, we refer to the two $0^+$ poles as $D_0^\ast(2100)$ and $D_0^\ast(2450)$, respectively. Their heavy-quark spin-symmetry partners are the $1^+$ S-wave $D_1(2250)$ and $D_1(2550)$~\cite{Albaladejo:2016lbb,Du:2017zvv}.

The $T$-matrix employed in this study is only valid up to $M_{X,0} \approx 2.6$~GeV. We continue the form factors for each channel separately at higher invariant masses:
\begin{align}
    f_X(q^2, M_X^2) = f_X(q^2,M^2_{X,0})\frac{M^2_{X,0}}{2M_X^2}\left(1 + \frac{M^{2n_X}_{X,0}}{M_X^{2n_X}}\right)~.\label{eq::cont}
\end{align}
This choice reproduces the expected asymptotic $1/M_X^2$ scaling and allows to match the first derivative at the matching point through an appropriate choice of $n_X$. In practice we find $n_{D\pi} = 0$, $n_{D\eta} = 8$, $n_{D_s K} = 5$. Comparing our results above $2.6$~GeV to the available $D\pi$ and $D_s K $ invariant mass spectra shows a fair description of the data, indicating that there are no sizeable deviations from the asymptotic scaling. In the case of the $D\eta$ channel, the decay rate had already dropped off considerably at $2.6$~GeV, indicating that our choice has little impact on the overall $B\rightarrow (D\eta)_S\ell\nu$ branching fraction.

With the continuation beyond $2.6$~GeV, the resulting branching fractions for the S-wave $D_s K$ and $D\eta$ contributions are
\begin{align}
    \mathcal{B}(B^+\rightarrow (D^-_s K^+)_S \ell^+\nu_\ell) &= (1.58 \pm 0.78)\times 10^{-4}~,\nonumber\\
    \mathcal{B}(B^+\rightarrow (D^0 \eta)_S \ell^+\nu_\ell) &= (0.56\pm 0.26)\times 10^{-4}~.
    \label{eq::dskdeta}
\end{align}
Both values are compatible, but more precise than the results reported in Ref.~\cite{Gustafson:2023lrz}, with the difference originating from the improved description of the S-wave. Approximately half of the observed $B^+\rightarrow D^-_s K^+ \ell^+\nu_\ell$ branching fraction can be explained by the S-wave component. Given the observed peak in the $D_s K$ invariant mass-spectrum around $2.6-2.8$~GeV~\cite{Belle:2012ccr}, the remainder likely comprises the $J^P=1^-$ resonances $D_1^\ast(2600)$ or $D_1^\ast(2760)$. The S-wave $D\eta$ component is small and can only account for approximately $1\%$ of the unobserved $\mathcal{B}(B^+\rightarrow D X \ell^+\nu_\ell)$ decays. However, given the presence of additional $D_s K$ modes, an additional P-wave $D\eta$ component of similar size to the S-wave should be expected.

Figure~\ref{fig:Fit} also shows the near-threshold region below $2.05$~GeV, which was not studied by the Belle analysis of $\mathcal{B}(B^+\rightarrow D^-\pi^+\ell^+\nu_\ell)$ decays. Although no experimental data is available to describe this region, our model predicts a clearly visible, non-vanishing P-wave component due to the subthreshold $D^\ast$. For this region, we find
\begin{align}
    \mathcal{B}(B^+\rightarrow D^- \pi^+ \ell^+\nu_\ell)\Big|_{\text{low-}M_{D\pi}} &= (2.11 \pm 0.18)\times 10^{-4}.\label{eq::lowmdpi}
\end{align}
As pointed out by Ref.~\cite{LeYaouanc:2021xcq}, this non-vanishing P-wave expected at threshold should be accounted for in both theoretical descriptions and future experimental analyses of $B \rightarrow D \pi \ell \nu$ decays.

We discuss additional results from the fit, including an updated determination of $\mathcal{B}(B \rightarrow D_{2}^{\ast}(2460) \ell \nu)$, in Appendix~\ref{sec::fit2}.

\section{Possible candidates for the semileptonic gap}\label{sec::candidates}
In the following, we discuss a non-exhaustive list of possible final states that will contribute in part to the semileptonic gap. No single decay mode is expected to close the remaining gap in its entirety, yet their sum might cover a sizeable portion. In addition, we aim to provide additional physics motivation to search for the presented modes. We include the little understood $D^{(\ast)}\pi\pi$ final states as their substructure might be used to predict additional final states.

Building on the results of Sec.~\ref{sec::swave}, we categorize the possible candidates by hadronic invariant mass. Aside from a low mass region, comprising the $D\pi$ threshold up to the narrow $D_1(2420)/D_2^\ast(2460)$ resonances, we sort the final states by the location of their threshold. This is motivated by the observation of threshold enhancements in many (quasi-)two-body channels, such as the aforementioned semileptonic $B\rightarrow (D\eta/D_s K)_S\ell\nu$ decays, $B^+\rightarrow p\bar{p}\ell^+\nu_\ell$ decays \cite{LHCb:2019cgl}, as well as nonleptonic decays like $B^+\rightarrow(\bar{\Lambda}^-_c p^+)_S\pi^+$ \cite{Belle:2004dmq,BaBar:2008get,LHCb:2024jqy}.

\begin{figure}[ht]
    \centering
    \includegraphics[width=0.48\textwidth]{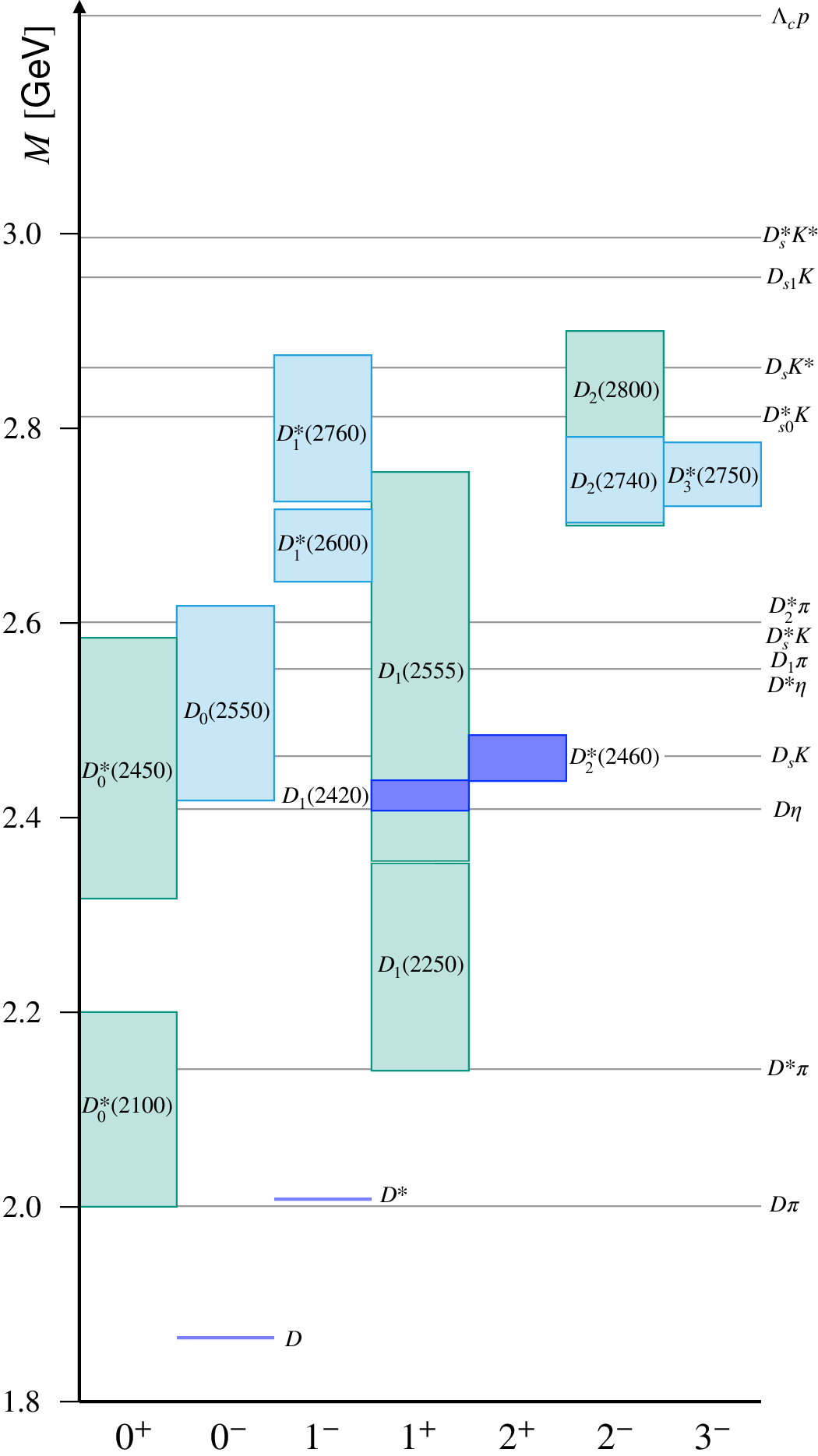}
    \caption{\label{fig:spectrum}
        Spectrum of non-strange charm mesons relative to (quasi-)two-body thresholds. The resonances in dark blue appear as narrow peaks in invariant mass distributions and are well-established, while the light-blue ones are broader and observed in angular analyses of $B\rightarrow D^{(\ast)}\pi\pi$ decays~\cite{LHCb:2016lxy,LHCb:2019juy}. Finally, the states in teal are theoretical predictions and we assume the masses and widths of Ref.~\cite{Du:2017zvv} for the $0^+$ and $1^+$ states, as well as Ref.~\cite{Godfrey:2015dva} for the broad $2^-$ resonance.
    }
\end{figure}

The known charm resonances, as well as the two $0^+$ and $1^+$ states are summarized together with the major (quasi-)two-body thresholds in Fig.~\ref{fig:spectrum}.

\subsection{Low mass region}
The first region of interest concerns the near-threshold region around the $D^\ast$, as well as the two narrow $D^{\ast\ast}$ resonances. All possible decay modes in this region will contribute towards the $D X \ell\nu$ gap and will exhibit kinematic distributions similar to those of narrow resonances. Consequently, they can contribute at high $q^2$ and lepton energy in the B rest frame $E_\ell^B$.

As the largest of all $b\rightarrow c\ell\nu$ transitions, the $B\rightarrow D^\ast\ell\nu$ branching fraction plays a crucial role in determining the size of the gap. Not only does it enter as a normalization mode in most inputs used in Sec.~\ref{sec::gapsize}, but it is also directly linked to a subset of $B\rightarrow D\pi\ell\nu$ decays, as demonstrated in Sec.~\ref{sec::swave}. This contribution is generally unaccounted for in determinations of $\mathcal{B}(B\rightarrow D^\ast\ell\nu)$. Additionally, mass windows used in determinations of $r^{(\ast)}_{\pi(\pi)}$ and $\mathcal{B}(B\rightarrow D^\ast\ell\nu)$ often differ for different experimental analyses~\cite{BaBar:2007nwi,BaBar:2007cke,Belle:2018ezy,Belle:2023bwv,Belle-II:2023okj}, potentially leading to inconsistencies.
While a detailed analysis of this problem is beyond the scope of this paper, it might not only shed light on the size of the semileptonic gap, but also the discrepancy in $|V_{cb}|$ determinations. However, several studies might shed light on the situation: a measurement of the $M_{D\pi}$-spectrum in $B^+\rightarrow D^-\pi^+\ell^+\nu_\ell$ decays including the threshold region would allow to directly test our prediction in Eq.~\eqref{eq::lowmdpi}. Furthermore, a direct measurement of $\mathcal{B}(B^0\rightarrow \bar{D}^0\pi^-\ell^+\nu_\ell)$ without subtracting the $D^\ast$ component could be used to confirm if
\begin{align}
    \frac{\mathcal{B}(B^0\rightarrow \bar{D}^0\pi^-\ell^+\nu_\ell)}{\mathcal{B}(B^0\rightarrow D^{\ast -}(\rightarrow \bar{D}^0\pi^-)\ell^+\nu_\ell)} =  1 + r_\pi~,
\end{align}
or uncover a mismatch in definitions. Lastly, measurements of the angular asymmetries described in Ref.~\cite{Du:2025beb} would not only provide the first direct evidence of the $D_{0}^{\ast}(2100)$, but also information on the $D^\ast$ tail.

The next set of decays of interest is $B\rightarrow D_1(2420)/D_2^\ast(2460)(\rightarrow D^{(\ast)}\pi\pi)\ell\nu$ decays. While the $D_1$ mode has been observed as the dominant component in $D\pi^+\pi^-$ decays~\cite{Belle:2022yzd}, it is unclear if the $D^\ast\pi\pi$ final states proceed through the narrow resonances and how to include the neutral pion modes, as discussed in Sec.~\ref{sec::gapsize}. As mentioned in Sec.~\ref{sec::gapsize}, nonleptonic measurements~\cite{Belle:2004bvv} favor the sequential emission scenario, for which HQET at leading power predicts~\cite{Falk:1992cx}:
\begin{align}
    \frac{\Gamma(D_1(2420)\rightarrow D^\ast\pi\pi)}{\Gamma(D_1(2420)\rightarrow D\pi\pi)}\Big|_{m_c\rightarrow\infty} &= \frac{1}{2}~,\nonumber\\
    \frac{\Gamma(D^\ast_2(2460)\rightarrow D^\ast\pi\pi)}{\Gamma(D_1(2420)\rightarrow D\pi\pi)}\Big|_{m_c\rightarrow\infty} &= \frac{3}{2}~.
\end{align}
This scenario predicts most of the observed $B\rightarrow D^\ast\pi\pi\ell\nu$ rate to be saturated by the $D_1(2420)$ and $D_2^\ast(2460)$. In practice, these relations will receive $1/m_c$ corrections and are modified by phase-space effects~\cite{Falk:1992cx}. Consequently, direct studies of the $D^{(\ast)}\pi\pi$ invariant mass spectrum and angular distributions are required to test the validity of this prediction.

Furthermore, the quark model calculation of Ref.~\cite{Godfrey:2015dva} predicts a sizeable $D^0_1(2420)\rightarrow D^0\gamma$, as well as a smaller $D_1^0(2420)/D_2^{\ast 0}(2460) \rightarrow D^{\ast 0}\gamma$  branching fraction. Although no uncertainties are quoted, the masses and widths of the two narrow states agree well with the predicted values. Consequently, the predictions for the partial widths should at least serve as a qualitative guideline. Both of these modes would appear as background in $R(D^{(\ast)})$ measurements. Combining the prediction of Ref.~\cite{Godfrey:2015dva} with the measured $B^+\rightarrow \bar{D}_1^0(2420)\ell^+\nu_\ell$ decay rates leads to an expectation of $\mathcal{B}(B^+\rightarrow \bar{D}^0_1(2420)(\rightarrow \bar{D}^0\gamma)\ell^+\nu_\ell$ of $\mathcal{O}(10^{-4})$, i.e. several percent of the missing $B^+\rightarrow DX\ell^+\nu_\ell$ components, placing it within reach of an analysis using the current Belle II data set.

\subsection{\texorpdfstring{$2.5-2.8$~}{2.5-2.8} GeV}
The second region of interest starts above the $D\eta$ and $D_sK$ threshold, where most of the observed $B\rightarrow D_s^{(\ast)}K\ell\nu$ decays are concentrated~\cite{Belle:2012ccr} and a sizeable fraction of $B\rightarrow D\pi\pi\ell\nu$ decays not stemming from the $D_1$ resonance are situated. In addition to the $0^+$ and $1^+$ S-wave states predicted in this region ~\cite{Albaladejo:2016lbb,Du:2017zvv}, five $D$-meson resonances have been observed in $D^{(\ast)}\pi$ final states: $D_0(2550)$, $D_1^\ast(2600)$, $D_2(2740)$, $D_3^\ast(2750)$, $D_1^\ast(2760)$ ~\cite{ParticleDataGroup:2024cfk}. The first two states can be attributed to the first radial excitations of the $D$ and $D^\ast$ mesons, the $2S$ doublet, while the latter three are good candidates to form the $1D$ quartett~\cite{Godfrey:2015dva}. From the quark model perspective, one $2^-$ state in this region, the $D_2(2800)$ is yet to be detected~\cite{Godfrey:2015dva}. The measured invariant mass spectra in $B\rightarrow D^{(\ast)}\pi\ell\nu$ do not reveal any clear peaks in this region, indicating that either the $D\pi$ and $D^\ast\pi$ final states are suppressed, or none of the five states can contribute substantially to the semileptonic gap through other modes.

The potentially largest contribution to the semileptonic gap from this mass-range originates from $B\rightarrow D^{(\ast)}\pi\pi\pi\ell\nu$ decays. With increasing pion multiplicity, combinatorial backgrounds become more challenging for experimental searches, consequently the most promising strategy is to search for decays involving intermediate narrow resonances.
In this mass range, three-pion contributions can originate from intermediate $\eta$ and $\omega$ mesons. While the $\eta$ contribution is likely small, as previously discussed, the $B\rightarrow D^{(\ast)}\omega\ell\nu$ decays are unconstrained. Given the narrow width of the $\omega$ and excellent understanding of the three-pion Dalitz plot~\cite{WASA-at-COSY:2016hfo}, combinatorial backgrounds can be effectively suppressed, rendering this mode ideal for a first direct search for $B\rightarrow D^{(\ast)}\pi\pi\pi\ell\nu$ decays.
In addition, a more careful study of decays with two final-state pions can provide indirect evidence for three-pion modes. Most prominent among them are the sequential decays $D_1^\ast(2760)\rightarrow D_1(2420)\pi$ and $D_2(2800)\rightarrow D_2^\ast(2460)\pi$, with subsequent $D_1(2420)/D_2^\ast(2460)\rightarrow D^{(\ast)}\pi$ decays~\cite{Godfrey:2015dva}. For example, in case a sizeable $D_1^\ast(2760)\rightarrow D_1(2420)(\rightarrow D^\ast\pi)\pi$ component is observed in $B\rightarrow D^{\ast}\pi\pi\ell\nu$ decays, this would directly imply a sizeable three-pion contribution through the large $D_1(2420)\rightarrow D\pi\pi$ decay rate. Similarly should the $D_2^\ast(2460)\rightarrow D^\ast\pi\pi$ decay be observed, the $D_2^\ast(2460)(\rightarrow D^{(\ast)}\pi)\pi$ final state would provide evidence for $B\rightarrow D^{\ast}\pi\pi\pi\ell\nu$ decays.

Finally, while $B\rightarrow D_s^{(\ast)}K\ell\nu$ decays have been observed and are taken into account in estimates of the size of the gap, improved studies of these final states have high physics potential. Furthermore, $B^0\rightarrow D_s^{(\ast)}K_L\ell^+\nu_\ell$ decays lead to similar experimental signatures as $B^0_s\rightarrow D^{(\ast)}_s\tau^+(\rightarrow \ell^+\bar{\nu}_\tau\nu_\ell)\nu_\tau$ decays. Most importantly however, the $D_s^{(\ast)}K$ threshold region is sensitive to the pole locations of the $D_0^\ast(2450)$ and $D_1(2550)$, enabling direct tests of their existence. To this end, it would be of great interest to study hadronic invariant mass spectra, as well as helicity angle distributions, separately for the $D_s$ and the $D_s^\ast$ modes. The angular distributions will allow to disentangle contributions of different partial waves and test the relative size of S-wave or $D_1^\ast$ contributions.

\subsection{\texorpdfstring{$2.8-3.1$~}{2.8-3.1} GeV}
Just below $3$~GeV several (quasi)-two-body decay channels of interest become kinematically available: $D_s^{(\ast)} K^\ast$, $D^\ast_{s0}(2317) K$, $D_{s1}(2460) K$, $D^{(\ast)} f_0(980)$, $D^{(\ast)} a_0(980)$, $D^{(\ast)}\eta^\prime$. The first three would contribute to the $D_s X \ell\nu$ gap, while the others contribute to the $D X \ell\nu$ gap. If the semileptonic gap is indeed in a large part due to $D_s$ mesons, the $D_s^{(\ast)} K^\ast$, $D^\ast_{s0}(2317) K$, $D_{s1}(2460) K$ final states need to contribute significantly to the missing fraction.

Decays into $D_s K^\ast$, $D^\ast_{s0}(2317) K$, $D f_0(980)$, as well as $D a_0(980)$ open up near $2.8$~GeV and would be kinematically favored, if proceeding through S-waves. Studies of the $D^{\ast}\pi$ final state in nonleptonic decays reveal non-trivial structures in the $0^-$ and $1^+$ partial waves in this region~\cite{LHCb:2019juy}. They can not be described by simple Breit-Wigner lineshapes and thus point to an interplay with the nearby thresholds.

Large $D^{(\ast)} f_0(980)/a_0(980)$ are particularly interesting, as they would result in $D^{(\ast)}K\bar{K}$ final states that remain to be observed experimentally. In particular, these mode would induce $B\rightarrow D^{(\ast)}K_L K_L\ell\nu$ decays, which would perfectly mimic $B\rightarrow D^{(\ast)}\tau(\rightarrow \ell\nu\nu)\nu$, impacting the interpretation of $R(D^{(\ast)})$ measurements. Given the smaller $D X \ell\nu$ gap, as well as the correlation with $K^+K^-$ and $K_S K_S$ final states, the impact on current measurements is expected to be small. Direct searches should take into account the $K^+K^-$ and $K^+ K_S$ final states, as their combination allows the separation of the $f_0$ and $a_0$ sub-components. Alternatively, the $a_0(980)$ component can be constrained through searches in the $\eta\pi$ decay channel.

Lastly, the $D^{(\ast)}\eta^\prime$ mode could be a sizeable component due to the large singlet component. It was observed in the $0^+$ $K\pi$ sector that the $K\eta^\prime$ channel has a larger impact on the inelasticity than the $K\eta$ channel~\cite{Jamin:2001zq} and a similar pattern could be realized in the charm sector. However, a direct search for $B\rightarrow D^{(\ast)}\eta^\prime\ell\nu$ decays poses a formidable task due to the complex decay channels of the $\eta^\prime$. 

\subsection{High mass region}
The region above $3.1$~GeV is of particular interest for semileptonic decays into baryonic final states. Although the first semileptonic decay involving baryons, the Cabibbo-suppressed $B^+\rightarrow p^+\bar{p}^-\ell^+\nu_\ell$ decay, has been observed by Belle~\cite{Belle:2013uqr} and confirmed by LHCb~\cite{LHCb:2019cgl}, only upper bounds exist for similar $b\rightarrow c$ transitions.
The most promising mode to search for in this region is the lightest baryonic final state, $B^+\rightarrow\Lambda_c^-p^+\ell^+\nu_\ell$. In addition to the BaBar limit on $B\rightarrow \Lambda_cX\ell\nu$ decays in Eq.~\eqref{eq::lcX}, CLEO set a limit on~\cite{CLEO:1997qbb}
\begin{align}
    \frac{\mathcal{B}(B^+\rightarrow \Lambda_c^-p^+e^+\nu_e)}{\mathcal{B}(B\rightarrow\Lambda_c^-p^+X)} < 0.04 \, ,
\end{align}
at the $90\%$ confidence level. The Review of Particle Physics to \cite{ParticleDataGroup:2024cfk} converts this result to
\begin{align}
    \mathcal{B}(B^+\rightarrow \Lambda_c^-p^+e^+\nu_e) < 8 \times 10^{-4}~.
\end{align}
Besides providing a clean channel to study the interaction of charm baryons with nucleons, a measurement of the $B^+\rightarrow \Lambda^-_c p^+\ell^+\nu_\ell$ branching fraction would directly provide the $B^0\rightarrow \Lambda^-_c n\ell^+\nu_\ell$ branching fraction. These decays could appear as a background in future studies of $\Lambda_b\rightarrow\Lambda_c\tau/\ell\nu$ with leptonic $\tau$ decays.

Furthermore, searches for $B\rightarrow \Sigma_c(\rightarrow \Lambda_c\pi) p \ell\nu$, $B\rightarrow \Lambda_c(2595)/\Lambda_c(2625)(\rightarrow\Lambda_c\pi\pi) p \ell\nu$ and $B\rightarrow \Xi_c^{(')}\Lambda\ell\nu$ should be feasible at Belle II, but likely require more data.

Besides baryonic final states, $B\rightarrow D_{s1}(2536)/D_{s2}^\ast(2573) K\ell\nu$ decays are allowed in this region, resulting in a mixture of $D^{(\ast)}_s K X$ and $D^{(\ast)} K \bar{K}$ final states. Direct searches for $B\rightarrow D_{s1}(2536)/D_{s2}^\ast(2573) K\ell\nu$ will allow to constrain the $D^{(\ast)} K_L K_L$ mode.

Finally, given the available phase-space, multi-pion final states will be ubiquitous in this region.

\section{Recommendations}\label{sec::Recommendations}
In this section, we provide recommendations for the treatment of the gap in event generation.
Furthermore, we propose several direct searches for gap modes of particular interest.

\subsection{\texorpdfstring{$B\rightarrow X_c\ell\nu$}{B→Xclν} modeling}
The main weakness of the treatment of the semileptonic gap by the experimental collaborations is the focus on one particular mode. This potentially leads to overestimated backgrounds in certain analyses, while underestimating possible backgrounds in others. To overcome this limitation, we propose to simulate the semileptonic gap with a mixture of plausible modes with loosely constrained normalization, allowing analyses to asses the impact of different final states.

We propose to include the following unobserved modes:
\begin{itemize}
    \item A small component comprising $D_1^0\rightarrow D^0\gamma$ decays, with a branching ratio of at most $10\%$.
    \item $B\rightarrow D\eta\ell\nu$ decays proceeding through the $D_0^\ast(2450)$ at the predicted rate together with a similarly sized contribution from $B\rightarrow D^{\ast}\eta\ell\nu$ decays proceeding through the $D_1(2550)$. An additional component proceeding through either the $D_1^\ast(2600)$ or $D_1^\ast(2760)$ could also be included, the size of which not exceeding twice the predicted $S$-wave rate.
    \item A pair of $0^-$ and $1^-$ resonances with a masses around $2.8$~GeV that decay to $D_{s0}^\ast(2317) K$ and $D a_0(980)$, as well as $D_{s1}^\ast(2460) K$ and $D^\ast a_0(980)$, respectively.\footnote{Modeling the $f_0(980)$ mode is more complex due to the strong $\pi\pi-K\bar{K}$ coupled-channel effects.}
    \item A pair of $0^+$ and a $1^+$ resonances with masses around $3$~GeV, decaying through the S-wave to $D_s^\ast K^\ast$ and $D_s K^\ast$, respectively.
    \item $B\rightarrow \Lambda_c p/n\ell\nu$ S-wave decays featuring a threshold enhancement, which is most easily achieved through the addition of a $0^-$ resonance at threshold. An additional contribution could also be included by allowing the same resonance to decay into $D_{s1}(2536)/D^\ast_{s2}(2573)K$.
\end{itemize}

In addition to including explicit gap contributions, the known $B\rightarrow D_s^{(\ast)}K\ell\nu$ and $B\rightarrow D^{(\ast)}\pi\pi\ell\nu$ decays should also be simulated.
\begin{itemize}
    \item Approximately half of $B\rightarrow D_s^{(\ast)}K\ell\nu$ decays should be simulated through a mixture of subthreshold $D_0^\ast(2450)$ and $D_1(2550)$ resonances for the $D_s$ and $D_s^\ast$ final states, respectively. The remainder could proceed through either the $D_1^\ast(2600)$ or $D_1^\ast(2760)$ resonances. Most importantly however, $B^0\rightarrow D_s K \ell\nu_\ell$ decays must also be considered in simulated samples.
    \item The measured $B\rightarrow D^\ast\pi\pi\ell\nu$ decay rate should be distributed over several modes. Part of it can be attributed to the $D_1(2420)/D_2^\ast(2460)$ resonances, while the remainder should proceed through the sequential decays $D^\ast_1(2760)\rightarrow D_1(\rightarrow D^\ast\pi)\pi$ and $D_2(2740)\rightarrow D_2^\ast(\rightarrow D^\ast\pi)\pi$. Here, the $D_2(2740)$ mode should not oversaturate the measured $D\pi\pi$ final state.\footnote{After subtraction of $D_1(2420)\rightarrow D\pi\pi$ decays.} Together, these decays already cover up to a quarter of the $DX\ell\nu$ gap through three-pion modes.
\end{itemize}

This “cocktail” provides a plethora of $B \rightarrow X_{c} \ell \nu$ final states, including $D^{(\ast)}\pi\pi\pi$, $D\gamma$, $D^{(\ast)}K\bar{K}$, $D_s^{(\ast)}K^{(\ast)}(\pi/\gamma)$, $\Lambda_c p$, $\Lambda_c n$, and $D^{(\ast)}\eta(\pi)$, resulting in a robust treatment of unmeasured modes that could contribute over a range of invariant masses and affect specific experimental analyses in distinct manners.
If simulated with the EvtGen event generator~\cite{Lange:2001uf}, all of the above modes must be corrected for missing or incorrect phase-space factors to obtain realistic distributions~\cite{Herren:2026pbh}.

\subsection{Experimental searches}
In addition to precision measurements of the observed channels, as well as possible semi-inclusive studies of $B\rightarrow D/D_s/\Lambda_c X\ell\nu$ decays, direct searches for possible gap modes are indispensable. In the following, we discuss some of the most promising channels:
\begin{itemize}
    \item Studies of $B\rightarrow D_s^{(\ast)} K^{(\ast)} \ell\nu$ decays should be the highest priority, given the large size of the $D_s/\Lambda_c$ gap. A new measurement of the $B\rightarrow D_s^{(\ast)} K\ell\nu$ branching fraction could improve our understanding of the only two known $B\rightarrow D_s X\ell\nu$ modes, while the inclusion of the $K^\ast$ resonance allows to test the existence of $D^{(\ast)}_s$ contributions beyond the ground state. Furthermore, improved $D_s^{(\ast)}K$ invariant mass spectra or measurements of the $D_s^{(\ast)}$ helicity angle spectra near the threshold could provide evidence for the existence of the $D_0^\ast(2450)$ and $D_1(2550)$ resonances. As an improvement to the existing analyses, future measurements should also take into account $B^0\rightarrow D_s^{(\ast)}K_S\ell\nu$ decays.
    \item A search of $D_1^0(2420)\rightarrow D^0\gamma$ decays is possible through the study of continuum $e^+ e^-\rightarrow c\bar{c}$ events at Belle II. By following a similar strategy to the analysis of the recent measurement of $\mathcal{B}(D_{s0}^\ast(2317)\rightarrow D_s^\ast\gamma) / \mathcal{B}(D_{s0}^\ast(2317)\rightarrow D_s\pi^0)$~\cite{Belle-II:2025dzk}, the ratios
    \begin{align}\frac{\mathcal{B}(D^0_{1}(2420)\rightarrow D^0\gamma)}{ \mathcal{B}(D_{1}^0(2420)\rightarrow D^{\ast +}(\rightarrow D^0 \pi^+)\pi^-)}~,\nonumber\end{align} or \begin{align}\frac{\mathcal{B}(D^0_{1}(2420)\rightarrow D^0\gamma)}{\mathcal{B}(D_{1}^0(2420)\rightarrow D^{\ast 0}(\rightarrow D^0 \gamma)\pi^0)}~,\nonumber\end{align} could be considered. For both ratios, systematic errors stemming from the production cross section and efficiencies of the $D^0$ decay channels cancel. While challenging, this measurement benefits from the large $D_1(2420)$ production cross-section of more than $30\,\text{pb}$ near the $\Upsilon(4S)$~\cite{CLEO:1994unc}.
    \item $B^+\rightarrow \Lambda_c^- p^+ \ell^+\nu_\ell$ decays provide the most simple decay signature involving baryons. A study at Belle II using hadronic tagging would provide a clean environment with manageable levels of combinatorial background or continuum processes. While efficiencies in hadronic tagged analyses are low, $\Lambda_c$ decay modes beyond the dominant $p^- K^+\pi^-$ channel could also be included. Most promising among them are two-body modes, such as $\bar{\Lambda}\pi^-$ and $p^-K_S$, but also modes with larger branching fractions, such as $p^-K_S\pi^+\pi^-$, $\bar{\Lambda}\pi^-\pi^0$ and $\bar{\Lambda}\pi^-\pi^+\pi^-$. Given the threshold enhancement present $B^+\rightarrow p^+\bar{p}^-\ell^+\nu_\ell$ and $B^+\rightarrow (\Lambda^-_c p^+)_S\pi^+$, a similar threshold enhancement is expected in $B^+\rightarrow \Lambda^-_c p^+\ell^+\nu_\ell$ decays and searches should be optimized for this scenario.
    \item Searches for $B\rightarrow D^{(\ast)} K^+ K^- \ell\nu$ decays are possible by following similar strategies to previous measurements of the $B\rightarrow D^{(\ast)} \pi^+ \pi^- \ell\nu$ branching fractions~\cite{BaBar:2015zkb,Belle:2022yzd}. Performing a hadronic tagged analysis and normalizing to the corresponding $B\rightarrow D^{(\ast)}\ell\nu$ yields provides a promising avenue at Belle II.
    \item Given the small branching fraction predicted for S-wave $B\rightarrow D^{(\ast)}\eta\ell\nu$ decays, observation with the current Belle II dataset might not be possible and a search at LHCb is complicated due to the neutral $\eta$ final states. However, even an upper limit of $\mathcal{O}(10^{-4})$ would reinforce our prediction, provide constraints on higher partial waves and improved understanding of the dynamics of this mode.
\end{itemize}

\section{Conclusion \texorpdfstring{\&}{and} Outlook}\label{sec::Outlook}
In this work, we discuss the current status of exclusive branching fraction measurements of semileptonic decays containing charm mesons in the final state. We use the available experimental data to investigate the underlying sources of the known difference between the sum of exclusive measurements and the inclusive $B\rightarrow X_c \ell\nu$ determination. We find that the remaining “gap” comprising unmeasured components in the total inclusive rates amounts to $1.8\%$ of all $B$ decays per light lepton flavor, or $17\%$ of all semileptonic decays. In addition, existing measurements suggest that more than half of the unknown final states contain $D_s$ mesons or $\Lambda_c$ baryons.

Our analysis of the semileptonic gap is limited to information on branching fractions, but could be augmented by kinematic information on inclusive and exclusive decays as in Ref.~\cite{Bernlochner:2014dca}. However, care needs to be taken, since several inclusive analyses were previously performed as sum-over-exclusive analyses. Additionally, the selection criteria of existing measurements of kinematic moments are optimized for $B\rightarrow D X\ell\nu$ decays, since $D_s$ or $\Lambda_c$ gap components are usually not considered. Furthermore, the required modeling of the kinematic spectra of the known $D^{\ast\ast}$ resonances relies on limited experimental information.

Additionally, we provide updated branching fraction predictions and improved descriptions for S-wave $B\rightarrow (D\eta/D_s K)_S\ell\nu$ decays; contributions of marginal size to the total semileptonic gap. The resulting branching fractions for the S-wave $D_s K$ and $D\eta$ contributions are
\begin{align}
    \mathcal{B}(B^+\rightarrow (D^-_s K^+)_S \ell^+\nu_\ell) &= (1.58 \pm 0.78)\times 10^{-4}~,\nonumber\\
    \mathcal{B}(B^+\rightarrow (D^0 \eta)_S \ell^+\nu_\ell) &= (0.56\pm 0.26)\times 10^{-4}~. \nonumber
\end{align}
Both values are compatible, but more precise than the results reported in Ref.~\cite{Gustafson:2023lrz}, with the difference originating from the improved description of the S-wave. While, the S-wave $D\eta$ component is small and can only account for approximately $1\%$ of the unobserved $\mathcal{B}(B^+\rightarrow D X \ell^+\nu_\ell)$ decays, an additional P-wave $D\eta$ component of similar size to the S-wave should be expected.

We discuss plausible candidates that could contribute to the semileptonic gap. While no single decay mode is expected to close the remaining gap in its entirety, their sum might constitute a sizeable portion. We provide recommendations for the treatment of the gap in Monte Carlo simulations, comprising a plethora of possible final states, which will allow future experimental measurements to assess systematic uncertainties due to the modelling of the unmeasured $B\rightarrow X_c \ell\nu$ modes in a robust manner.

Finally, we propose direct searches for the most promising candidates, all of which could be conducted in the near future at Belle II. Searches for these candidate modes will not only close the knowledge gap in semileptonic $B\rightarrow X_c\ell\nu$ decays, but also provide invaluable insights into the spectroscopy of excited charm mesons.

\section*{Acknowledgments}
We thank Biljana Mitreska for explanations on the treatment of the semileptonic gap within LHCb and Florian Bernlochner for pointing out Ref.~\cite{Bernlochner:2014dca}.

This research is supported in part by the Swiss
National Science Foundation (SNF) under contract 200021-212729. RvT is supported by the German Research Foundation (DFG) Walter-Benjamin Grant No. 545582477.

\section*{Data availability}
We will provide a \verb|Python| module implementing the S-wave line shapes and form factors, the relevant fitted parameters, as well as examples reproducing Figs.~\ref{fig:Swaves} and~\ref{fig:Bs} in a notebook. 

\appendix
\section{Different input for \texorpdfstring{$B\rightarrow D^{(\ast)}\ell\nu$}{B→D(*)lν} branching fractions}\label{sec::further_fits}
To investigate the dependence of our results on the $B\rightarrow D^{(\ast)}\ell\nu$ branching fractions, we use the results of Ref.~\cite{Jung:2026ewj}, which corrects different measurements for the d’Agostini bias~\cite{DAgostini:1993arp} and removes overlooked inconsistencies in the treatment of older results.
With the branching fractions from Tbl. 4 of Ref.~\cite{Jung:2026ewj}, we obtain:
\begin{align}
    \mathcal{B}(B^+\rightarrow D_s/\Lambda_c\,X \ell^+\nu_\ell)\Big|_\mathrm{gap} &= (1.25\pm 0.36)\%~,\nonumber\\
    \mathcal{B}(B^+\rightarrow D X \ell^+\nu_\ell)\Big|_\mathrm{gap} &= (0.45\pm 0.24)\%~,
\end{align}
for the $B^+$ decays and
\begin{align}
    \mathcal{B}(B^0\rightarrow D_s/\Lambda_c\,X \ell^+\nu_\ell)\Big|_\mathrm{gap} &= (0.66\pm 0.38)\%~,\nonumber\\
    \mathcal{B}(B^0\rightarrow D X \ell^+\nu_\ell)\Big|_\mathrm{gap} &= (0.99\pm 0.35)\%~,
\end{align}
for $B^0$ decays. The combined gap sizes are
\begin{align}
    \mathcal{B}(B^+\rightarrow X_c \ell^+\nu_\ell)\Big|_\mathrm{gap} &= (1.75 \pm 0.31)\%~,\nonumber\\
    \mathcal{B}(B^0\rightarrow X_c \ell^+\nu_\ell)\Big|_\mathrm{gap} &= (1.62 \pm 0.23)\%~.
\end{align}

In the combined fit scenario, the branching fractions are given by
\begin{align}
    \mathcal{B}(B\rightarrow D_s/\Lambda_c\,X \ell\nu)\Big|_\mathrm{gap} &= (1.00\pm 0.28)\%\nonumber\\
    \mathcal{B}(B\rightarrow DX \ell\nu)\phantom{\Big|_\mathrm{gap}} &= (9.57\pm 0.26)\%\nonumber\\
    \mathcal{B}(B^+\rightarrow D X \ell^+\nu_\ell)\Big|_\mathrm{gap} &= (0.57\pm 0.24)\%\nonumber\\
    \mathcal{B}(B^0\rightarrow D X \ell^+\nu_\ell)\Big|_\mathrm{gap} &= (0.70\pm 0.27)\%~.
    \label{eq::fitresults2}
\end{align}

All results are well compatible with our main fit. The size of the $D_s/\Lambda_c$ gap component is reduced by $10\%$, but remains larger than the gap containing $D$ mesons.

\section{\texorpdfstring{$B_s$}{Bs} decays}\label{sec::bsdecays}
In addition to the $D\pi-D\eta-D_s K$ $T$-matrix, the results of Ref.~\cite{Liu:2012zya} allow to obtain the $D K-D_s\eta$ $T$-matrix. The latter is relevant for the description of $B_s\rightarrow (D K)_S\ell\nu$ decays, which are of relevance for determinations of the ratio of $B_s$ and $B_d$ production fractions~\cite{LHCb:2011leg,LHCb:2019fns} and sum-of-exclusive measurements of the hadronic mass moments in $B_s$ decays~\cite{DeCian:2023ezb} at LHCb.

\begin{figure}[ht]
    \centering
    \includegraphics[width=0.4825\textwidth]{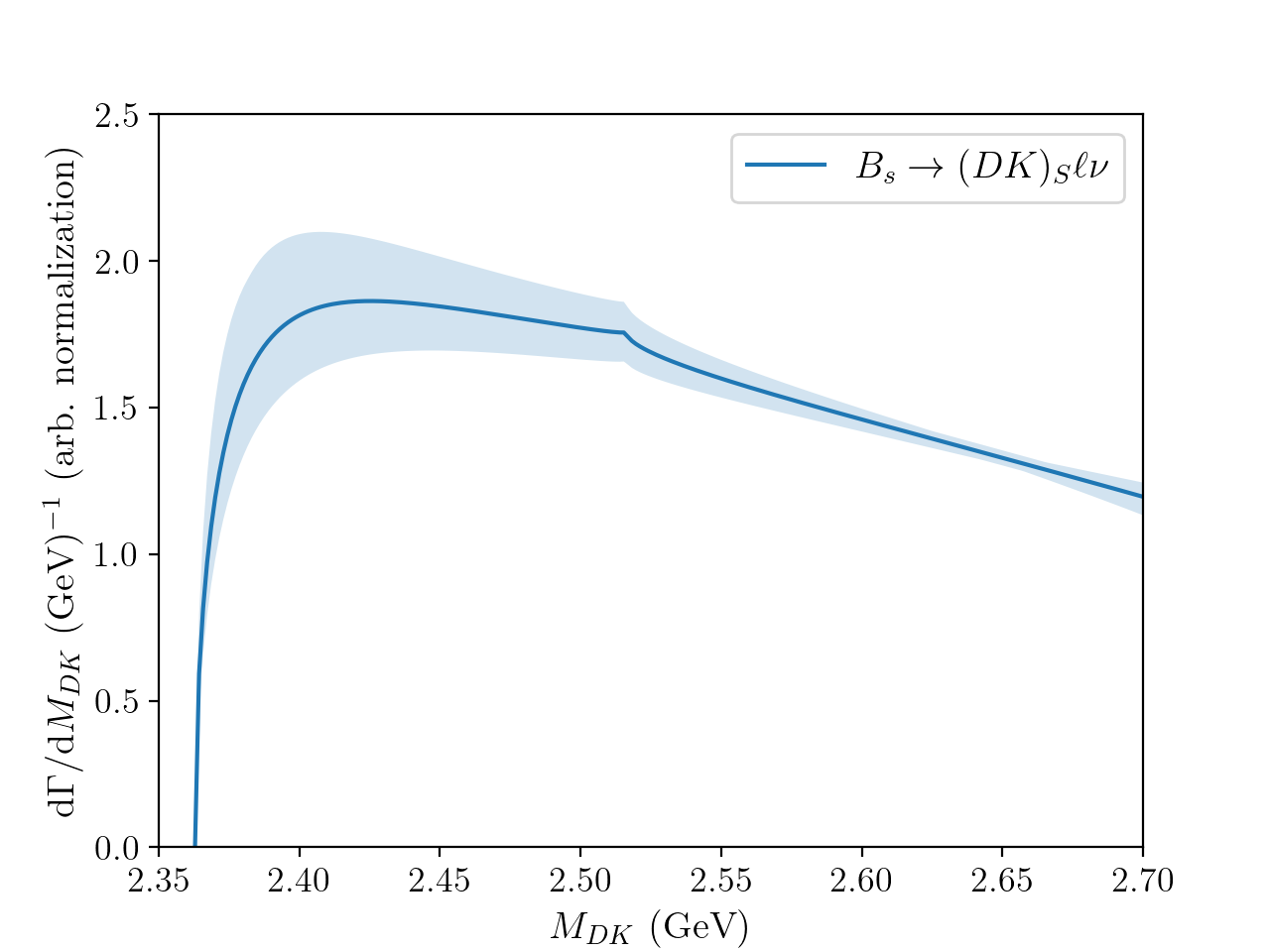}
    \caption{\label{fig:Bs}
        $B_s \rightarrow (D K)_S \ell\nu_\ell$ lineshape. The normalization is unknown and only shape uncertainties are shown.
    }
\end{figure}

In this case, the two form factors are given by
\renewcommand\arraystretch{1.2}
\begin{align}
    \begin{pmatrix}
    f_{DK}(q^2,M_X^2) \\ f_{D_s\eta}(q^2,M_X^2)
    \end{pmatrix}
    = T(M_X^2)
    \begin{pmatrix}
    1 \\ -1/\sqrt{6}\,
    \end{pmatrix} P(q^2)~.
\end{align}
\renewcommand\arraystretch{1.0}
The function $P(q^2)$ can be parametrized as in Eq.~\eqref{eq::swave_ff}. While the normalization and $q^2$-dependence are unknown, the $M_{DK}$ dependence itself is fully determined by the $T$-matrix up until $2.7$~GeV. Above $2.7$~GeV, we continue the form factors following Eq.~\eqref{eq::cont} with $n_{DK} = n_{D_s\eta} = 0$. The resulting $DK$ lineshape is shown in Fig.~\ref{fig:Bs} and clearly dominated by the subthreshold $D_{s0}^\ast(2317)$. A small cusp is visible at the $D_s\eta$ threshold.

\section{Additional fit results}\label{sec::fit2}
Compared to Ref.~\cite{Gustafson:2023lrz}, the branching fractions for the individual partial waves show sizeable shifts. First, the changes in the treatment of the energy-dependent width and the Blatt--Weisskopf damping factor in the description of the $D^\ast$ lead to a longer tail, effectively increasing the P-wave component. Second, with our improved treatment of the S-wave, the fit does not have as much freedom as in Ref.~\cite{Gustafson:2023lrz} to change its shape to accommodate a larger $D_2^\ast(2460)$ contribution.

For the total S-wave contribution, we obtain
\begin{align}
    \mathcal{B}(B^+\rightarrow (D^-\pi^+)_S\ell^+\nu_\ell) = (1.85\pm 0.41) \times 10^{-3}~.
\end{align}
The central value is a factor of $1.8$ larger than in Ref.~\cite{Gustafson:2023lrz} and the relative uncertainty amounts to $22\%$ instead of $25\%$. This change can be fully attributed to the sizeable contribution of the region between $2.4$~GeV and $2.5$~GeV, i.e. underneath the $D_2^\ast(2460)$, where the second pole is located. In Ref.~\cite{Gustafson:2023lrz} the S-wave had a negligible contribution in this region, since the fit had greater freedom to adjust the lineshape.

In the P-wave, the $D_1^\ast(2600)$ starts to contribute above $2.5$~GeV and consequently we refrain from including this region when quoting our result.
The result from threshold up to $M_{D\pi} = 2.5~\text{GeV}$ is given by
\begin{align}
    \mathcal{B}(B^+\rightarrow (D^-\pi^+)_P\ell^+\nu_\ell)\Big|_{M_{D\pi} <2.5~\text{GeV}}\nonumber\\  = (1.10\pm 0.10) \times 10^{-3}~.
\end{align}
This result is $20\%$ larger than in Ref.~\cite{Gustafson:2023lrz}, but accounting for uncertainties the change is at most at the $2\sigma$ level.

Finally, the $D_2^\ast(2460)$ contribution is $10\%$ smaller than in Ref.~\cite{Gustafson:2023lrz} and now in good agreement with the Belle result \cite{Belle:2022yzd}:
\begin{align}
    \mathcal{B}(B^+\rightarrow \bar{D}_2^{\ast0}(2460)(\rightarrow D^-\pi^+)\ell^+\nu_\ell)\nonumber\\ = (1.71\pm 0.15) \times 10^{-3}~.
\end{align}

The changes in our branching fractions clearly underline the complexity of extracting branching fractions of overlapping resonances and other contributions from the invariant mass spectrum alone. For reliable results either angular information~\cite{Du:2025beb} or precisely known lineshapes~\cite{Herren:2025cwv} are required.

\newpage
\bibliography{refs}

@article{Herren:2026pbh,
    author = "Herren, Florian and van Tonder, Raynette",
    title = "{On the simulated kinematic distributions of semileptonic $B$ decays}",
    eprint = "2602.18378",
    archivePrefix = "arXiv",
    primaryClass = "hep-ph",
    reportNumber = "ZU-TH 08/26",
    month = "2",
    year = "2026"
}

@article{Du:2025beb,
    author = "Du, M. -L. and Guo, F. -K. and Hanhart, C. and Herren, F. and Kubis, B. and van Tonder, R.",
    title = "{Discovering the $D_0^*(2100)$ in B semileptonic decays}",
    eprint = "2509.12133",
    archivePrefix = "arXiv",
    primaryClass = "hep-ph",
    reportNumber = "ZU-TH 57/25",
    doi = "10.1140/epjc/s10052-025-15035-7",
    journal = "Eur. Phys. J. C",
    volume = "85",
    number = "11",
    pages = "1289",
    year = "2025"
}

@article{Gustafson:2023lrz,
    author = "Gustafson, Erik J. and Herren, Florian and Van de Water, Ruth S. and van Tonder, Raynette and Wagman, Michael L.",
    title = "{Model-independent description of $B\rightarrow D\pi\ell\nu$ decays}",
    eprint = "2311.00864",
    archivePrefix = "arXiv",
    primaryClass = "hep-ph",
    reportNumber = "FERMILAB-PUB-23-662-T, ZU-TH 67/23",
    doi = "10.1103/PhysRevD.110.L091502",
    journal = "Phys. Rev. D",
    volume = "110",
    number = "9",
    pages = "L091502",
    year = "2024"
}

@article{Rudolph:2018rzl,
    author = "Rudolph, Matthew",
    title = "{An experimentalist{\textquoteright}s guide to the semileptonic bottom to charm branching fractions}",
    eprint = "1805.05659",
    archivePrefix = "arXiv",
    primaryClass = "hep-ph",
    doi = "10.1142/S0217751X18501762",
    journal = "Int. J. Mod. Phys. A",
    volume = "33",
    number = "32",
    pages = "1850176",
    year = "2018"
}

@article{Bernlochner:2012bc,
    author = "Bernlochner, Florian U. and Ligeti, Zoltan and Turczyk, Sascha",
    title = "{A Proposal to solve some puzzles in semileptonic B decays}",
    eprint = "1202.1834",
    archivePrefix = "arXiv",
    primaryClass = "hep-ph",
    doi = "10.1103/PhysRevD.85.094033",
    journal = "Phys. Rev. D",
    volume = "85",
    pages = "094033",
    year = "2012"
}

@article{Bernlochner:2014dca,
    author = {Bernlochner, Florian U. and Biedermann, Dustin and Lacker, Heiko and L{\"u}ck, Thomas},
    title = "{Constraints on exclusive branching fractions ${\mathcal {B}}_i(B^+\rightarrow X_c^il^+\nu )$ from moment measurements in inclusive $B\rightarrow X_cl\nu $ decays}",
    eprint = "1402.2849",
    archivePrefix = "arXiv",
    primaryClass = "hep-ph",
    doi = "10.1140/epjc/s10052-014-2914-3",
    journal = "Eur. Phys. J. C",
    volume = "74",
    number = "6",
    pages = "2914",
    year = "2014"
}

@article{LHCb:2018azb,
    author = "Aaij, Roel and others",
    collaboration = "LHCb",
    title = "{Measurement of the relative $B^{-} \!\rightarrow D^{0} / D^{*0} / D^{**0} \mu^{-} \overline{\nu}_\mu$ branching fractions using $B^{-}$ mesons from $\overline{B}{}_{s2}^{*0}$ decays}",
    eprint = "1807.10722",
    archivePrefix = "arXiv",
    primaryClass = "hep-ex",
    reportNumber = "LHCb-PAPER-2018-024, CERN-EP-2018-190",
    doi = "10.1103/PhysRevD.99.092009",
    journal = "Phys. Rev. D",
    volume = "99",
    number = "9",
    pages = "092009",
    year = "2019"
}

@article{BaBar:2007xlq,
    author = "Aubert, Bernard and others",
    collaboration = "BaBar",
    title = "{Measurement of the relative branching fractions of $\bar{B} \to$ D/D*/D** $\ell^{-} \bar{\nu}$( $\ell^{)}$ decays in events with a fully reconstructed $B$ meson}",
    eprint = "hep-ex/0703027",
    archivePrefix = "arXiv",
    reportNumber = "SLAC-PUB-12393, BABAR-PUB-07-012",
    doi = "10.1103/PhysRevD.76.051101",
    journal = "Phys. Rev. D",
    volume = "76",
    pages = "051101",
    year = "2007"
}

@article{HeavyFlavorAveragingGroupHFLAV:2024ctg,
    author = "Banerjee, Sw. and others",
    collaboration = "Heavy Flavor Averaging Group (HFLAV)",
    title = "{Averages of b-hadron, c-hadron, and $\tau$-lepton properties as of 2023}",
    eprint = "2411.18639",
    archivePrefix = "arXiv",
    primaryClass = "hep-ex",
    doi = "10.1103/x87q-tld5",
    journal = "Phys. Rev. D",
    volume = "113",
    number = "1",
    pages = "012008",
    year = "2026"
}

@article{Belle:2022yzd,
    author = "Meier, F. and others",
    collaboration = "Belle",
    title = "{First observation of $B\rightarrow \bar{D}_1 (\rightarrow \bar{D}\pi^+\pi^-)\ell^+\nu_\ell$ and measurement of the $B\rightarrow \bar{D}^{(\ast)} \pi\ell^+\nu_\ell$ and $B\rightarrow \bar{D}^{(\ast)} \pi^+\pi^-\ell^+\nu_\ell$ branching fractions with hadronic tagging at Belle}",
    eprint = "2211.09833",
    archivePrefix = "arXiv",
    primaryClass = "hep-ex",
    reportNumber = "Belle Preprint 2022-32, KEK Preprint 2022-44",
    doi = "10.1103/PhysRevD.107.092003",
    journal = "Phys. Rev. D",
    volume = "107",
    number = "9",
    pages = "092003",
    year = "2023"
}

@article{BaBar:2015zkb,
    author = "Lees, J. P. and others",
    collaboration = "BaBar",
    title = "{Observation of $\overline{B} \to D^{(*)} \pi^+\pi^- \ell^-\overline{\nu}$ decays in $e^+e^-$ collisions at the $\Upsilon(4S)$ resonance}",
    eprint = "1507.08303",
    archivePrefix = "arXiv",
    primaryClass = "hep-ex",
    reportNumber = "BABAR-PUB-15-007, SLAC-PUB-16337",
    doi = "10.1103/PhysRevLett.116.041801",
    journal = "Phys. Rev. Lett.",
    volume = "116",
    number = "4",
    pages = "041801",
    year = "2016"
}

@article{Belle:2012ccr,
    author = "Stypula, J. and others",
    collaboration = "Belle",
    title = "{Evidence for $B^-\to D_s^+ K^- \ell^-\bar{\nu}_\ell$ and search for $B^-\to D_s^{*+} K^- \ell^-\bar{\nu}_\ell$}",
    eprint = "1207.6244",
    archivePrefix = "arXiv",
    primaryClass = "hep-ex",
    doi = "10.1103/PhysRevD.86.072007",
    journal = "Phys. Rev. D",
    volume = "86",
    pages = "072007",
    year = "2012"
}

@article{BaBar:2010ner,
    author = "del Amo Sanchez, P. and others",
    collaboration = "BaBar",
    title = "{Observation of the Decay $B^{-} \rightarrow D_{s}^{(*)+} K^{-} \ell^{-} \bar{\nu}_{\ell}$}",
    eprint = "1012.4158",
    archivePrefix = "arXiv",
    primaryClass = "hep-ex",
    reportNumber = "SLAC-PUB-14245, BABAR-PUB-10-010",
    doi = "10.1103/PhysRevLett.107.041804",
    journal = "Phys. Rev. Lett.",
    volume = "107",
    pages = "041804",
    year = "2011"
}

@article{BaBar:2007ddh,
    author = "Aubert, Bernard and others",
    collaboration = "BaBar",
    title = "{A Measurement of the branching fractions of exclusive $\bar{B} \to D^{(*)}$ ($\pi$) $\ell^{-} \bar{\nu}$( $\ell^{)}$ decays in events with a fully reconstructed $B$ meson}",
    eprint = "0712.3503",
    archivePrefix = "arXiv",
    primaryClass = "hep-ex",
    reportNumber = "SLAC-PUB-13056, BABAR-PUB-07-071",
    doi = "10.1103/PhysRevLett.100.151802",
    journal = "Phys. Rev. Lett.",
    volume = "100",
    pages = "151802",
    year = "2008"
}

@article{ParticleDataGroup:2024cfk,
    author = "Navas, S. and others",
    collaboration = "Particle Data Group",
    title = "{Review of particle physics}",
    doi = "10.1103/PhysRevD.110.030001",
    journal = "Phys. Rev. D",
    volume = "110",
    number = "3",
    pages = "030001",
    year = "2024"
}

@article{Belle:2004bvv,
    author = "Abe, Kazuo and others",
    collaboration = "Belle",
    title = "{Observation of the D(1)(2420) ---{\ensuremath{>}} D pi+ pi- decays}",
    eprint = "hep-ex/0410091",
    archivePrefix = "arXiv",
    doi = "10.1103/PhysRevLett.94.221805",
    journal = "Phys. Rev. Lett.",
    volume = "94",
    pages = "221805",
    year = "2005"
}

@article{Charles:2004jd,
    author = "Charles, J. and Hocker, Andreas and Lacker, H. and Laplace, S. and Le Diberder, F. R. and Malcles, J. and Ocariz, J. and Pivk, M. and Roos, L.",
    collaboration = "CKMfitter Group",
    title = "{CP violation and the CKM matrix: Assessing the impact of the asymmetric $B$ factories}",
    eprint = "hep-ph/0406184",
    archivePrefix = "arXiv",
    reportNumber = "CPT-2004-P-030, LAL-04-21, LAPP-EXP-2004-01, LPNHE-2004-01",
    doi = "10.1140/epjc/s2005-02169-1",
    journal = "Eur. Phys. J. C",
    volume = "41",
    number = "1",
    pages = "1--131",
    year = "2005"
}

@article{UTfit:2022hsi,
    author = "Bona, Marcella and others",
    collaboration = "UTfit",
    title = "{New UTfit Analysis of the Unitarity Triangle in the Cabibbo-Kobayashi-Maskawa scheme}",
    eprint = "2212.03894",
    archivePrefix = "arXiv",
    primaryClass = "hep-ph",
    reportNumber = "YITP-SB-2022-40",
    doi = "10.1007/s12210-023-01137-5",
    journal = "Rend. Lincei Sci. Fis. Nat.",
    volume = "34",
    pages = "37--57",
    year = "2023"
}

@article{Finauri:2023kte,
    author = "Finauri, Gael and Gambino, Paolo",
    title = "{The q$^{2}$ moments in inclusive semileptonic B decays}",
    eprint = "2310.20324",
    archivePrefix = "arXiv",
    primaryClass = "hep-ph",
    reportNumber = "TUM-HEP 1477/23",
    doi = "10.1007/JHEP02(2024)206",
    journal = "JHEP",
    volume = "02",
    pages = "206",
    year = "2024"
}

@article{Higson:2018cwj,
    author = "Higson, Edward and Handley, Will and Hobson, Michael and Lasenby, Anthony",
    title = "{Dynamic nested sampling: an improved algorithm for parameter estimation and evidence calculation}",
    eprint = "1704.03459",
    archivePrefix = "arXiv",
    primaryClass = "stat.CO",
    doi = "10.1007/s11222-018-9844-0",
    journal = "Stat. Comput.",
    volume = "29",
    number = "5",
    pages = "891--913",
    year = "2018"
}

@article{Speagle:2019ivv,
    author = "Speagle, Joshua S.",
    title = "{dynesty: a dynamic nested sampling package for estimating Bayesian posteriors and evidences}",
    eprint = "1904.02180",
    archivePrefix = "arXiv",
    primaryClass = "astro-ph.IM",
    doi = "10.1093/mnras/staa278",
    journal = "Mon. Not. Roy. Astron. Soc.",
    volume = "493",
    number = "3",
    pages = "3132--3158",
    year = "2020"
}

@article{ARGUS:1993hhz,
    author = "Albrecht, H. and others",
    collaboration = "ARGUS",
    title = "{Search for rare B meson decays into D(s)+ mesons}",
    reportNumber = "DESY-93-054",
    doi = "10.1007/BF01650427",
    journal = "Z. Phys. C",
    volume = "60",
    pages = "11--18",
    year = "1993"
}

@article{CLEO:1997qbb,
    author = "Bonvicini, G. and others",
    collaboration = "CLEO",
    title = "{Study of semileptonic decays of B mesons to charmed baryons}",
    eprint = "hep-ex/9712008",
    archivePrefix = "arXiv",
    reportNumber = "SLAC-PUB-9757, CLNS-97-1519, CLEO-97-26",
    doi = "10.1103/PhysRevD.57.6604",
    journal = "Phys. Rev. D",
    volume = "57",
    pages = "6604--6608",
    year = "1998"
}

@article{BaBar:2011ijz,
    author = "Lees, J. P. and others",
    collaboration = "BaBar",
    title = "{Search for $\bar{B} --> \Lambda_c+ X l- \nu_l$ Decays in Events With a Fully Reconstructed $B$ Meson}",
    eprint = "1110.6005",
    archivePrefix = "arXiv",
    primaryClass = "hep-ex",
    reportNumber = "BABAR-PUB-11-011, SLAC-PUB-14659",
    doi = "10.1103/PhysRevD.85.011102",
    journal = "Phys. Rev. D",
    volume = "85",
    pages = "011102",
    year = "2012"
}

@article{Albaladejo:2016lbb,
    author = "Albaladejo, Miguel and Fernandez-Soler, Pedro and Guo, Feng-Kun and Nieves, Juan",
    title = "{Two-pole structure of the $D^\ast_0(2400)$}",
    eprint = "1610.06727",
    archivePrefix = "arXiv",
    primaryClass = "hep-ph",
    doi = "10.1016/j.physletb.2017.02.036",
    journal = "Phys. Lett. B",
    volume = "767",
    pages = "465--469",
    year = "2017"
}

@article{Du:2017zvv,
    author = "Du, Meng-Lin and Albaladejo, Miguel and Fern{\'a}ndez-Soler, Pedro and Guo, Feng-Kun and Hanhart, Christoph and Mei{\ss}ner, Ulf-G. and Nieves, Juan and Yao, De-Liang",
    title = "{Towards a new paradigm for heavy-light meson spectroscopy}",
    eprint = "1712.07957",
    archivePrefix = "arXiv",
    primaryClass = "hep-ph",
    doi = "10.1103/PhysRevD.98.094018",
    journal = "Phys. Rev. D",
    volume = "98",
    number = "9",
    pages = "094018",
    year = "2018"
}

@article{Belle-II:2025dzk,
    author = "Abumusabh, M. and others",
    collaboration = "Belle-II",
    title = "{Observation of the radiative decay $D_s (2317)^+ \to D_s^* γ$}",
    eprint = "2510.27174",
    archivePrefix = "arXiv",
    primaryClass = "hep-ex",
    reportNumber = "Belle II Preprint {\#}2025-026? KEK Preprint {\#}2025-28",
    month = "10",
    year = "2025"
}

@article{Belle-II:2025pye,
    author = "Abumusabh, M. and others",
    collaboration = "Belle-II",
    title = "{Measurement of inclusive B{\textrightarrow}Xu{\ensuremath{\ell}}{\ensuremath{\nu}} partial branching fractions and |Vub| at Belle II}",
    eprint = "2512.08056",
    archivePrefix = "arXiv",
    primaryClass = "hep-ex",
    doi = "10.1103/59ws-zxbt",
    journal = "Phys. Rev. D",
    volume = "113",
    number = "3",
    pages = "032004",
    year = "2026"
}

@article{Belle-II:2025yjp,
    author = "Adachi, I. and others",
    collaboration = "Belle-II",
    title = "{Test of lepton flavor universality with measurements of R(D+) and R(D*+) using semileptonic B tagging at the Belle II experiment}",
    eprint = "2504.11220",
    archivePrefix = "arXiv",
    primaryClass = "hep-ex",
    reportNumber = "Belle II Preprint 2025-011, KEK Preprint 2025-9",
    doi = "10.1103/fmn3-h8fy",
    journal = "Phys. Rev. D",
    volume = "112",
    number = "3",
    pages = "032010",
    year = "2025"
}

@article{Belle-II:2024ami,
    author = "Adachi, I. and others",
    collaboration = "Belle-II",
    title = "{Test of lepton flavor universality with a measurement of R(D*) using hadronic B tagging at the Belle II experiment}",
    eprint = "2401.02840",
    archivePrefix = "arXiv",
    primaryClass = "hep-ex",
    reportNumber = "BELLE2-PUB-PH-2024-001 https://docs.belle2.org/record/4057/ ; Belle
  II Preprint 2024-001; KEK Preprint 2023-47",
    doi = "10.1103/PhysRevD.110.072020",
    journal = "Phys. Rev. D",
    volume = "110",
    number = "7",
    pages = "072020",
    year = "2024"
}

@article{Belle:2021idw,
    author = "van Tonder, R. and others",
    collaboration = "Belle",
    title = "{Measurements of $q^2$ Moments of Inclusive $B \rightarrow X_c \ell^+ \nu_{\ell}$ Decays with Hadronic Tagging}",
    eprint = "2109.01685",
    archivePrefix = "arXiv",
    primaryClass = "hep-ex",
    reportNumber = "Belle Preprint 2021-18, KEK Preprint 2021-22",
    doi = "10.1103/PhysRevD.104.112011",
    journal = "Phys. Rev. D",
    volume = "104",
    number = "11",
    pages = "112011",
    year = "2021"
}

@article{Belle:2021eni,
    author = "Cao, L. and others",
    collaboration = "Belle",
    title = "{Measurements of Partial Branching Fractions of Inclusive $B \to X_u \, \ell^+\, \nu_{\ell}$ Decays with Hadronic Tagging}",
    eprint = "2102.00020",
    archivePrefix = "arXiv",
    primaryClass = "hep-ex",
    reportNumber = "Belle Preprint 2020-22, KEK Preprint 2020-39",
    doi = "10.1103/PhysRevD.104.012008",
    journal = "Phys. Rev. D",
    volume = "104",
    number = "1",
    pages = "012008",
    year = "2021"
}

@article{Belle-II:2023aih,
    author = "Adachi, I. and others",
    collaboration = "Belle-II",
    title = "{First Measurement of R(X{\ensuremath{\tau}}/{\ensuremath{\ell}}) as an Inclusive Test of the b{\textrightarrow}c{\ensuremath{\tau}}{\ensuremath{\nu}} Anomaly}",
    eprint = "2311.07248",
    archivePrefix = "arXiv",
    primaryClass = "hep-ex",
    reportNumber = "Belle II Preprint 2023-016, KEK Preprint 2023-34",
    doi = "10.1103/PhysRevLett.132.211804",
    journal = "Phys. Rev. Lett.",
    volume = "132",
    number = "21",
    pages = "211804",
    year = "2024"
}

@article{LHCb:2023zxo,
    author = "Aaij, Roel and others",
    collaboration = "LHCb",
    title = "{Measurement of the ratios of branching fractions $\mathcal{R}(D^{*})$ and $\mathcal{R}(D^{0})$}",
    eprint = "2302.02886",
    archivePrefix = "arXiv",
    primaryClass = "hep-ex",
    reportNumber = "LHCb-PAPER-2022-039, CERN-EP-2022-284",
    doi = "10.1103/PhysRevLett.131.111802",
    journal = "Phys. Rev. Lett.",
    volume = "131",
    pages = "111802",
    year = "2023"
}

@article{LHCb:2023uiv,
    author = "Aaij, Roel and others",
    collaboration = "LHCb",
    title = "{Test of lepton flavor universality using $B_0 \rightarrow D^{\ast -} \tau^+ \nu_\tau$ decays with hadronic \ensuremath{\tau} channels}",
    eprint = "2305.01463",
    archivePrefix = "arXiv",
    primaryClass = "hep-ex",
    reportNumber = "LHCb-PAPER-2022-052, CERN-EP-2023-062",
    doi = "10.1103/PhysRevD.108.012018",
    journal = "Phys. Rev. D",
    volume = "108",
    number = "1",
    pages = "012018",
    year = "2023"
}

@article{Godfrey:2015dva,
    author = "Godfrey, Stephen and Moats, Kenneth",
    title = "{Properties of Excited Charm and Charm-Strange Mesons}",
    eprint = "1510.08305",
    archivePrefix = "arXiv",
    primaryClass = "hep-ph",
    doi = "10.1103/PhysRevD.93.034035",
    journal = "Phys. Rev. D",
    volume = "93",
    number = "3",
    pages = "034035",
    year = "2016"
}

@article{Falk:1992cx,
    author = "Falk, Adam F. and Luke, Michael E.",
    title = "{Strong decays of excited heavy mesons in chiral perturbation theory}",
    eprint = "hep-ph/9206241",
    archivePrefix = "arXiv",
    reportNumber = "SLAC-PUB-5812, UCSD-PTH-92-14",
    doi = "10.1016/0370-2693(92)90618-E",
    journal = "Phys. Lett. B",
    volume = "292",
    pages = "119--127",
    year = "1992"
}

@article{LHCb:2019juy,
    author = "Aaij, Roel and others",
    collaboration = "LHCb",
    title = "{Determination of quantum numbers for several excited charmed mesons observed in $B^- \to D^{*+} \pi^- \pi^-$ decays}",
    eprint = "1911.05957",
    archivePrefix = "arXiv",
    primaryClass = "hep-ex",
    reportNumber = "CERN-EP-2019-201, LHCb-PAPER-2019-027",
    doi = "10.1103/PhysRevD.101.032005",
    journal = "Phys. Rev. D",
    volume = "101",
    number = "3",
    pages = "032005",
    year = "2020"
}

@article{LHCb:2016lxy,
    author = "Aaij, Roel and others",
    collaboration = "LHCb",
    title = "{Amplitude analysis of $B^{-} \to D^{+} \pi^{-} \pi^{-}$ decays}",
    eprint = "1608.01289",
    archivePrefix = "arXiv",
    primaryClass = "hep-ex",
    reportNumber = "CERN-EP-2016-184, LHCB-PAPER-2016-026",
    doi = "10.1103/PhysRevD.94.072001",
    journal = "Phys. Rev. D",
    volume = "94",
    number = "7",
    pages = "072001",
    year = "2016"
}

@article{Bernlochner:2016bci,
    author = "Bernlochner, Florian U. and Ligeti, Zoltan",
    title = "{Semileptonic $B_{(s)}$ decays to excited charmed mesons with $e,\mu,\tau$ and searching for new physics with $R(D^{**})$}",
    eprint = "1606.09300",
    archivePrefix = "arXiv",
    primaryClass = "hep-ph",
    doi = "10.1103/PhysRevD.95.014022",
    journal = "Phys. Rev. D",
    volume = "95",
    number = "1",
    pages = "014022",
    year = "2017"
}

@phdthesis{Lueck2013Determination,
author = {L\"uck, Thomas},
title = {Determination of the CKM-matrix element $|V_{ub}|$ from the electron energy spectrum measured in inclusive $B\rightarrow X_u e \nu$ decay with the BABAR detector},
school = {Humboldt-Universit\"at zu Berlin, Mathematisch-Naturwissenschaftliche Fakult\"at I},
year = {2013},
doi = {http://dx.doi.org/10.18452/16718}
}

@article{Jung:2026ewj,
    author = "Jung, Martin and Schacht, Stefan",
    title = "{$\bar B\to D^{(*)}\ell\bar ν$ Branching Ratios and Evidence for Isospin Breaking in $Υ(4S)$ Decays}",
    eprint = "2604.08391",
    archivePrefix = "arXiv",
    primaryClass = "hep-ph",
    reportNumber = "IPPP/26/30",
    month = "4",
    year = "2026"
}

@article{VonHippel:1972fg,
    author = "Von Hippel, F. and Quigg, C.",
    title = "{Centrifugal-barrier effects in resonance partial decay widths, shapes, and production amplitudes}",
    doi = "10.1103/PhysRevD.5.624",
    journal = "Phys. Rev. D",
    volume = "5",
    pages = "624--638",
    year = "1972"
}

@article{Belle:2007uwr,
    author = "Liventsev, D. and others",
    collaboration = "Belle",
    title = "{Study of B --\ensuremath{>} D**lnu with full reconstruction tagging}",
    eprint = "0711.3252",
    archivePrefix = "arXiv",
    primaryClass = "hep-ex",
    reportNumber = "BELLE-PREPRINT-2007-46, KEK-PREPRINT-2007-55",
    doi = "10.1103/PhysRevD.77.091503",
    journal = "Phys. Rev. D",
    volume = "77",
    pages = "091503",
    year = "2008"
}

@article{FermilabLattice:2021cdg,
    author = "Bazavov, A. and others",
    collaboration = "Fermilab Lattice, MILC, Fermilab Lattice, MILC",
    title = "{Semileptonic form factors for $B\rightarrow D^*\ell \nu $ at nonzero recoil from $2+1$-flavor lattice QCD: Fermilab Lattice~and~MILC~Collaborations}",
    eprint = "2105.14019",
    archivePrefix = "arXiv",
    primaryClass = "hep-lat",
    reportNumber = "FERMILAB-PUB-21-261-T~, FERMILAB-PUB-21/261-T",
    doi = "10.1140/epjc/s10052-022-10984-9",
    journal = "Eur. Phys. J. C",
    volume = "82",
    number = "12",
    pages = "1141",
    year = "2022",
    note = "[Erratum: Eur. Phys. J. C \textbf{83}, 21 (2023)]"
}

@article{Herren:2025cwv,
    author = "Herren, Florian and Kubis, Bastian and van Tonder, Raynette",
    title = "{Model-independent parametrization of $B\rightarrow\pi\pi\ell\nu$ decays}",
    eprint = "2502.20960",
    archivePrefix = "arXiv",
    primaryClass = "hep-ph",
    reportNumber = "EOS-2025-02, ZU-TH 04/25",
    doi = "10.1103/8pm6-9xzq",
    journal = "Phys. Rev. D",
    volume = "112",
    number = "1",
    pages = "014037",
    year = "2025"
}

@article{Liu:2012zya,
    author = "Liu, Liuming and Orginos, Kostas and Guo, Feng-Kun and Hanhart, Christoph and Mei{\ss}ner, Ulf-G.",
    title = "{Interactions of charmed mesons with light pseudoscalar mesons from lattice QCD and implications on the nature of the $D_{s0}^*(2317)$}",
    eprint = "1208.4535",
    archivePrefix = "arXiv",
    primaryClass = "hep-lat",
    reportNumber = "JLAB-THY-12-1599",
    doi = "10.1103/PhysRevD.87.014508",
    journal = "Phys. Rev. D",
    volume = "87",
    number = "1",
    pages = "014508",
    year = "2013"
}

@article{BaBar:2007nwi,
    author = "Aubert, Bernard and others",
    collaboration = "BaBar",
    title = "{Measurement of the Decay $B^{-} \rightarrow D^{\ast0} e^{-} \bar{\nu}_e$}",
    eprint = "0712.3493",
    archivePrefix = "arXiv",
    primaryClass = "hep-ex",
    reportNumber = "SLAC-PUB-13035, BABAR-PUB-07-070",
    doi = "10.1103/PhysRevLett.100.231803",
    journal = "Phys. Rev. Lett.",
    volume = "100",
    pages = "231803",
    year = "2008"
}

@article{BaBar:2007cke,
    author = "Aubert, Bernard and others",
    collaboration = "BaBar",
    title = "{Determination of the form-factors for the decay $B^0 \to D^{*-} \ell^{+} \nu_{l}$ and of the CKM matrix element $|V_{cb}|$}",
    eprint = "0705.4008",
    archivePrefix = "arXiv",
    primaryClass = "hep-ex",
    reportNumber = "SLAC-PUB-12511, BABAR-PUB-07-008",
    doi = "10.1103/PhysRevD.77.032002",
    journal = "Phys. Rev. D",
    volume = "77",
    pages = "032002",
    year = "2008"
}

@article{Belle-II:2023okj,
    author = "Adachi, I. and others",
    collaboration = "Belle-II",
    title = "{Determination of |Vcb| using B{\textasciimacron}0{\textrightarrow}D*+{\ensuremath{\ell}}{\ensuremath{-}}{\ensuremath{\nu}}{\textasciimacron}{\ensuremath{\ell}} decays with Belle II}",
    eprint = "2310.01170",
    archivePrefix = "arXiv",
    primaryClass = "hep-ex",
    reportNumber = "Belle II Preprint 2023-014, KEK Preprint 2023-28",
    doi = "10.1103/PhysRevD.108.092013",
    journal = "Phys. Rev. D",
    volume = "108",
    number = "9",
    pages = "092013",
    year = "2023"
}

@article{Belle:2018ezy,
    author = "Waheed, E. and others",
    collaboration = "Belle",
    title = "{Measurement of the CKM matrix element $|V_{cb}|$ from $B^0\to D^{*-}\ell^ {+} \nu_\ell$ at Belle}",
    eprint = "1809.03290",
    archivePrefix = "arXiv",
    primaryClass = "hep-ex",
    doi = "10.1103/PhysRevD.100.052007",
    journal = "Phys. Rev. D",
    volume = "100",
    number = "5",
    pages = "052007",
    year = "2019",
    note = "[Erratum: Phys.Rev.D 103, 079901 (2021)]"
}

@article{Lepage:2001ym,
    author = "Lepage, G. P. and Clark, B. and Davies, C. T. H. and Hornbostel, K. and Mackenzie, P. B. and Morningstar, C. and Trottier, H.",
    editor = "Muller-Preussker, M. and Bietenholz, Wolfgang and Jansen, K. and Jegerlehner, F. and Montvay, I. and Schierholz, G. and Sommer, R. and Wolff, U.",
    collaboration = "HPQCD",
    title = "{Constrained curve fitting}",
    eprint = "hep-lat/0110175",
    archivePrefix = "arXiv",
    reportNumber = "FERMILAB-CONF-01-485-T",
    doi = "10.1016/S0920-5632(01)01638-3",
    journal = "Nucl. Phys. B Proc. Suppl.",
    volume = "106",
    pages = "12--20",
    year = "2002"
}

@article{Hornbostel:2011hu,
    author = "Hornbostel, K. and Lepage, G. P. and Davies, C. T. H. and Dowdall, R. J. and Na, H. and Shigemitsu, J.",
    title = "{Fast Fits for Lattice QCD Correlators}",
    eprint = "1111.1363",
    archivePrefix = "arXiv",
    primaryClass = "hep-lat",
    doi = "10.1103/PhysRevD.85.031504",
    journal = "Phys. Rev. D",
    volume = "85",
    pages = "031504",
    year = "2012"
}

@article{LHCb:2019cgl,
    author = "Aaij, Roel and others",
    collaboration = "LHCb",
    title = "{Observation of the semileptonic decay $B^{+}\to p\overline{p}\mu^{+}\nu_{\mu}$}",
    eprint = "1911.08187",
    archivePrefix = "arXiv",
    primaryClass = "hep-ex",
    reportNumber = "LHCb-PAPER-2019-034, CERN-EP-2019-236",
    doi = "10.1007/JHEP03(2020)146",
    journal = "JHEP",
    volume = "03",
    pages = "146",
    year = "2020"
}

@article{Belle:2004dmq,
    author = "Gabyshev, N. and others",
    collaboration = "Belle",
    title = "{Study of decay mechanisms in B- ---{\ensuremath{>}} Lambda/c+ anti-p pi- decays and observation of low-mass structure in the Lambda/c+ anti-p system}",
    eprint = "hep-ex/0409005",
    archivePrefix = "arXiv",
    reportNumber = "BELLE-PREPRINT-2006-31, KEK-PREPRINT-2006-48",
    doi = "10.1103/PhysRevLett.97.242001",
    journal = "Phys. Rev. Lett.",
    volume = "97",
    pages = "242001",
    year = "2006"
}

@article{LHCb:2024jqy,
    author = "Aaij, Roel and others",
    collaboration = "LHCb",
    title = "{Study of b-hadron decays to $ {\Lambda}_c^{+}{h}^{-}{h}^{\prime -} $ final states}",
    eprint = "2405.12688",
    archivePrefix = "arXiv",
    primaryClass = "hep-ex",
    reportNumber = "CERN-EP-2024-116, LHCb-PAPER-2024-013",
    doi = "10.1007/JHEP08(2024)132",
    journal = "JHEP",
    volume = "08",
    pages = "132",
    year = "2024"
}

@article{BaBar:2008get,
    author = "Aubert, Bernard and others",
    collaboration = "BaBar",
    title = "{Measurements of B(anti-B0 ---{\ensuremath{>}} Lambda(c)+ anti-p) and B(B- ---{\ensuremath{>}} Lambda(c)+ anti-p pi-) and Studies of Lambda(c)+ pi- Resonances}",
    eprint = "0807.4974",
    archivePrefix = "arXiv",
    primaryClass = "hep-ex",
    reportNumber = "BABAR-PUB-08-016, SLAC-PUB-13341",
    doi = "10.1103/PhysRevD.78.112003",
    journal = "Phys. Rev. D",
    volume = "78",
    pages = "112003",
    year = "2008"
}

@article{Lange:2001uf,
    author = "Lange, D. J.",
    editor = "Erhan, S. and Schlein, P. and Rozen, Y.",
    title = "{The EvtGen particle decay simulation package}",
    doi = "10.1016/S0168-9002(01)00089-4",
    journal = "Nucl. Instrum. Meth. A",
    volume = "462",
    pages = "152--155",
    year = "2001"
}

@article{LeYaouanc:2018zyo,
    author = "Le Yaouanc, Alain and Leroy, Jean-Pierre and Roudeau, Patrick",
    title = "{Large off-shell effects in the $\bar{D}^*$ contribution to $B \to \bar{D} \pi \pi$ and $B \to \bar{D} \pi \bar{\ell} \nu_{\ell}$ decays.}",
    eprint = "1806.09853",
    archivePrefix = "arXiv",
    primaryClass = "hep-ph",
    reportNumber = "LPT-Orsay-18-51, LAL 18-011, LPT-ORSAY-18-51, LAL-18-011",
    doi = "10.1103/PhysRevD.99.073010",
    journal = "Phys. Rev. D",
    volume = "99",
    number = "7",
    pages = "073010",
    year = "2019"
}

@article{LeYaouanc:2021xcq,
    author = "Le Yaouanc, Alain and Leroy, Jean-Pierre and Roudeau, Patrick",
    title = "{Model for nonleptonic and semileptonic decays by ${\overline{B}}^{0}{\rightarrow}{D}^{**}$ transitions with $\mathrm{BR}(j=1/2){\ll}\mathrm{BR}(j=3/2)$ using the Leibovich-Ligeti-Stewart-Wise schem}",
    eprint = "2102.11608",
    archivePrefix = "arXiv",
    primaryClass = "hep-ph",
    doi = "10.1103/PhysRevD.105.013004",
    journal = "Phys. Rev. D",
    volume = "105",
    number = "1",
    pages = "013004",
    year = "2022"
}

@article{WASA-at-COSY:2016hfo,
    author = "Adlarson, P. and others",
    collaboration = "WASA-at-COSY",
    title = "{Measurement of the $\omega \to \pi^+ \pi^- \pi^0$ Dalitz plot distribution}",
    eprint = "1610.02187",
    archivePrefix = "arXiv",
    primaryClass = "nucl-ex",
    doi = "10.1016/j.physletb.2017.03.050",
    journal = "Phys. Lett. B",
    volume = "770",
    pages = "418--425",
    year = "2017"
}

@article{Jamin:2001zq,
    author = "Jamin, Matthias and Oller, Jose Antonio and Pich, Antonio",
    title = "{Strangeness changing scalar form-factors}",
    eprint = "hep-ph/0110193",
    archivePrefix = "arXiv",
    reportNumber = "IFIC-01-26, FTUV-01-1015, HD-THEP-01-10, FZ-IKP-TH-01-16",
    doi = "10.1016/S0550-3213(01)00605-8",
    journal = "Nucl. Phys. B",
    volume = "622",
    pages = "279--308",
    year = "2002"
}

@article{Becirevic:2012te,
    author = "Becirevic, Damir and Le Yaouanc, Alain and Oliver, Luis and Raynal, Jean-Claude and Roudeau, Patrick and Serrano, Justine",
    title = "{Proposal to study Bs{\textrightarrow}D{\textasciimacron}sJ transitions}",
    eprint = "1206.5869",
    archivePrefix = "arXiv",
    primaryClass = "hep-ph",
    reportNumber = "LAL-12-212, LPT-12-58",
    doi = "10.1103/PhysRevD.87.054007",
    journal = "Phys. Rev. D",
    volume = "87",
    number = "5",
    pages = "054007",
    year = "2013"
}

@article{Belle:2013uqr,
    author = "Tien, K. -J. and others",
    collaboration = "Belle",
    title = "{Evidence for semileptonic $B^- \to p\bar{p}l^-\bar{\nu}_l$ decays}",
    eprint = "1306.3353",
    archivePrefix = "arXiv",
    primaryClass = "hep-ex",
    reportNumber = "BELLE-PREPRINT-2013-10, KEK-PREPRINT-2013-10",
    doi = "10.1103/PhysRevD.89.011101",
    journal = "Phys. Rev. D",
    volume = "89",
    number = "1",
    pages = "011101",
    year = "2014"
}

@article{CLEO:1994unc,
    author = "Avery, P. and others",
    collaboration = "CLEO",
    title = "{Production and decay of D$_1$(2420)$^0$ and D$_2^*$(2460)$^0$}",
    eprint = "hep-ph/9403359",
    archivePrefix = "arXiv",
    reportNumber = "CLNS-94-1280, CLEO-94-10, CLNS94-1280",
    doi = "10.1016/0370-2693(94)90968-7",
    journal = "Phys. Lett. B",
    volume = "331",
    pages = "236--244",
    year = "1994",
    note = "[Erratum: Phys.Lett.B 342, 453--453 (1995)]"
}

@article{LHCb:2019fns,
    author = "Aaij, Roel and others",
    collaboration = "LHCb",
    title = "{Measurement of $b$ hadron fractions in 13 TeV $pp$ collisions}",
    eprint = "1902.06794",
    archivePrefix = "arXiv",
    primaryClass = "hep-ex",
    reportNumber = "CERN-EP-2019-016, LHCb-PAPER-2018-050",
    doi = "10.1103/PhysRevD.100.031102",
    journal = "Phys. Rev. D",
    volume = "100",
    number = "3",
    pages = "031102",
    year = "2019"
}

@article{DAgostini:1993arp,
    author = "D'Agostini, G.",
    title = "{On the use of the covariance matrix to fit correlated data}",
    reportNumber = "DESY-93-175",
    doi = "10.1016/0168-9002(94)90719-6",
    journal = "Nucl. Instrum. Meth. A",
    volume = "346",
    pages = "306--311",
    year = "1994"
}

@article{Meissner:2020khl,
    author = "Mei{\ss}ner, Ulf-G.",
    title = "{Two-pole structures in QCD: Facts, not fantasy!}",
    eprint = "2005.06909",
    archivePrefix = "arXiv",
    primaryClass = "hep-ph",
    doi = "10.3390/sym12060981",
    journal = "Symmetry",
    volume = "12",
    number = "6",
    pages = "981",
    year = "2020"
}

@article{Belle:2023bwv,
    author = "Prim, M. T. and others",
    collaboration = "Belle",
    title = "{Measurement of differential distributions of B{\textrightarrow}D*{\ensuremath{\ell}}{\ensuremath{\nu}}{\textasciimacron}{\ensuremath{\ell}} and implications on |Vcb|}",
    eprint = "2301.07529",
    archivePrefix = "arXiv",
    primaryClass = "hep-ex",
    reportNumber = "Belle Preprint 2022-34; KEK Preprint 2022-47",
    doi = "10.1103/PhysRevD.108.012002",
    journal = "Phys. Rev. D",
    volume = "108",
    number = "1",
    pages = "012002",
    year = "2023"
}

@article{Bolognani:2026yxj,
    author = "Bolognani, Carolina and Jung, Martin and Reboud, M{\'e}ril and Keri Vos, K.",
    title = "{Extracting production fractions of $b$ hadrons from exclusive semi-leptonic decays}",
    eprint = "2605.00181",
    archivePrefix = "arXiv",
    primaryClass = "hep-ph",
    reportNumber = "EOS-2026-02",
    month = "4",
    year = "2026"
}

@article{LHCb:2011leg,
    author = "Aaij, R. and others",
    collaboration = "LHCb",
    title = "{Measurement of $b$-hadron production fractions in $7~\rm{TeV}$ pp collisions}",
    eprint = "1111.2357",
    archivePrefix = "arXiv",
    primaryClass = "hep-ex",
    reportNumber = "CERN-PH-EP-2011-172, LHCB-PAPER-2011-018",
    doi = "10.1103/PhysRevD.85.032008",
    journal = "Phys. Rev. D",
    volume = "85",
    pages = "032008",
    year = "2012"
}

@article{DeCian:2023ezb,
    author = "De Cian, M. and Feliks, N. and Rotondo, M. and Keri Vos, K.",
    title = "{Inclusive semileptonic $ {B}_s^0 $ meson decays at the LHC via a sum-of-exclusive modes technique: possibilities and prospects}",
    eprint = "2312.05147",
    archivePrefix = "arXiv",
    primaryClass = "hep-ph",
    doi = "10.1007/JHEP06(2024)158",
    journal = "JHEP",
    volume = "06",
    pages = "158",
    year = "2024"
}

@article{EOSAuthors:2021xpv,
    author = "van Dyk, Danny and others",
    collaboration = "EOS Authors",
    title = "{EOS: a software for flavor physics phenomenology}",
    eprint = "2111.15428",
    archivePrefix = "arXiv",
    primaryClass = "hep-ph",
    reportNumber = "EOS-2021-04, TUM-HEP 1371/21, P3H-21-094, SI-HEP-2021-32",
    doi = "10.1140/epjc/s10052-022-10177-4",
    journal = "Eur. Phys. J. C",
    volume = "82",
    number = "6",
    pages = "569",
    year = "2022"
}

@software{the_eos_authors_collaboration_2025_17792609,
  author       = "van Dyk, Danny and others",
  collaboration = "EOS Authors",
  title        = {{EOS} Version 1.10.19},
  month        = dec,
  year         = 2025,
  publisher    = {Zenodo},
  version      = {v1.0.19},
  doi          = {10.5281/zenodo.17792609},
  url          = {https://doi.org/10.5281/zenodo.17792609},
  swhid        = {swh:1:dir:15380be12f8d8cbbcae967f3f24092d4036982e1
                   ;origin=https://doi.org/10.5281/zenodo.3376590;vis
                   it=swh:1:snp:47012bd580b65b25da243a25580331b307919
                   7f5;anchor=swh:1:rel:ff533a1f851aab5c39eca084e35bd
                   f688aa28e6c;path=eos-eos-37907e2
                  },
}

\end{document}